\documentclass{emulateapj}
\usepackage{amsmath}

\begin{document}
\def\GS{\text{GD-1 }}

\ifnum-1=-1
\title{Constraining the Milky Way potential with a 
6-D phase-space map of the \GS stellar stream
\footnotemark[*]}
\footnotetext[*]{Based on observations collected at the
Centro Astron\'omico Hispano Alem\'an (CAHA) at Calar Alto, operated jointly by
the Max-Planck Institut f\"ur Astronomie and the Instituto de Astrof\'isica de
Andaluc\'ia(CSIC).}
\else
\title{Constraining the Milky Way potential with a 
6-D phase-space map of the \GS stellar stream
\footnote{Based on observations collected at the
Centro Astron\'omico Hispano Alem\'an (CAHA) at Calar Alto, operated jointly by
the Max-Planck Institut f\"ur Astronomie and the Instituto de Astrof\'isica de
Andaluc\'ia(CSIC).}}
\fi
\author{Sergey E. Koposov,\altaffilmark{1,2,3},
	Hans-Walter Rix\altaffilmark{1},
	David W. Hogg\altaffilmark{1,4}
}
\altaffiltext{1}{Max Planck Institute for Astronomy, K\"{o}nigstuhl
17, 69117 Heidelberg, Germany\email{koposov@ast.cam.ac.uk}}
\altaffiltext{2}{Institute of Astronomy, Madingley Road, Cambridge CB3 0HA,
UK}
\altaffiltext{3}{Sternberg Astronomical Institute, Universitetskiy pr. 13,
119992 Moscow, Russia}
\altaffiltext{4}{Center for Cosmology and Particle Physics, Department of
Physics, New York University, 4 Washington Place, New York, NY 10003, USA}

\def\phione{\phi_1}
\def\phitwo{\phi_2}

\def\muone{\mu_{\phione}}
\def\mutwo{\mu_{\phitwo}}
\def\pphn{\phn\phn}
\def\ppphn{\phn\phn\phn}
\def\change#1{#1}

\begin{abstract}
The narrow GD-1 stream of stars, spanning 60~deg on the sky at a
distance of $\sim 10$~kpc from the Sun and $\sim 15$~kpc from the Galactic
center, is presumed to be debris from a tidally disrupted star cluster that
traces out a test-particle orbit in the Milky Way
halo. We combine SDSS photometry, USNO-B astrometry, and SDSS and
Calar Alto spectroscopy to construct a complete, empirical
6-dimensional phase-space map of the stream.  We find that an
eccentric orbit in a flattened isothermal potential describes this
phase-space map well. Even after marginalizing over the stream
orbital parameters and the distance from the Sun to the Galactic
center, the orbital fit to GD-1 places strong constraints on the
circular velocity at the Sun's radius $V_c=224\pm13$~km/s and total potential
flattening $q_\Phi=0.87^{+0.07}_{-0.04}$.  When we drop any informative priors
on $V_c$ the GD-1 constraint becomes $V_c=221\pm18$~km/s. Our 6-D map of GD-1
therefore
yields the best current constraint on $V_c$ and the only strong
constraint on $q_\Phi$ at Galactocentric radii near R$\sim15$~kpc. Much, if not
all, of the total potential flattening may be attributed to the mass
in the stellar disk, so the GD-1 constraints on the flattening of the
halo itself are weak: $q_{\Phi,halo}>0.89$ at 90\% confidence. The greatest
uncertainty in the 6-D map and the
orbital analysis stems from the photometric distances,
which will be obviated by Gaia.
\end{abstract}
\section{Introduction}

The Sloan Digital Sky Survey (SDSS) is an imaging and spectroscopy survey which
mapped quarter of the sky near the North Galactic Cap. The data have
proven extremely
useful for the understanding of the Milky Way halo. In addition to a large
list of MW
satellites \citep{belukurov07_quintet,koposov07_gc,irwin07,walsh07} several
extended stellar sub-structures in the MW halo have been found in the SDSS data,
such
as the tidal tail of the Palomar 5 globular cluster
\citep{odenkirchen01,grillmair06_pal5},
the Monoceros ring \citep{newberg02}, two northern tidal arms of the
disrupting Sagittarius galaxy \citep{belokurov06_field_of_streams}, the so called
``Orphan'' stream \citep{grillmair06_orphan,belokurov07_orphan}, the Aquila
overdensity \citep{belokurov07_aquila} and the very long thin stellar stream
called \GS \citep{grillmair06_gd}. \change{Recently \citet{grillmair09} claimed the
discovery of another 4 stellar streams.}
 Streams are presumed to be remnants of
tidally disrupted satellite galaxies and clusters. They
provide important insights into the history of accretion events and the physics
of Galaxy formation. The tidal debris from disrupted
satellites (clusters) spreads out in orbital phase on a path that is close to
the orbit of the progenitor. 
Streams tracing out orbits therefore provide opportunities to
constrain the Milky Way's gravitational potential.

After initial searches for tidal tails of globular clusters
\citep[e.g.][]{grillmair95} it was the extended Sagittarius tidal tail that
first made deriving such constraints practical \citep[see
e.g.][]{ibata01,helmi04,johnston05,law05}. 
However, the tidal tail of the Sagittarius galaxy is quite wide and 
contains a considerable mixture of different stellar orbits, making it complex
to model. For constraining
the gravitational potential, a stellar stream that is very thin but of large
angular extent, is ideal, because it permits precise orbital models.

The first studies
of globular cluster tidal debris only revealed short ($\lesssim$ 1$\degr$)
signs of tails, but in recent years with the advent of large
photometric surveys such as SDSS and 2MASS and significant advances in the
techniques used to find streams, significant progress has been made. The matched
filter technique \citep{odenkirchen01,rockosi02} has revealed the beautiful
tidal stream of Palomar~5.
Detailed analysis  of the Pal~5 stream, including kinematics
\citep{odenkirchen01,odenkirchen03,odenkirchen09},
have shown the promise of this approach, but also revealed that data over
more than 10$\degr$ on the sky are needed to place good constraints on the
potential.
\citet{grillmair06_orphan}, \citet{grillmair06_pal5,grillmair06_gd},
\citet{grillmair06_ngc5466} were
successful in the detection of very long stellar streams using this technique,
including the 63$\degr$ long stellar stream \GS. Besides the stream length
and the approximate distance, most of the properties of \GS
 were unknown. Since the stream is long but relatively
thin, with no apparent progenitor remnant, it was suggested that it arose from a
globular cluster. 
In this paper we make an attempt to determine all possible properties of the
\GS stream including distance, position on the sky, proper motion, and radial
velocity and try to constrain the Milky Way potential using that information.
This work goes in parallel with the work done by~\citet{willett09}, but we are
able to get a full 6-D phase space map of the stream and are
able to use that map to provide significant constraints on the MW
potential. See also~\citet{eyre09} for theoretical discussion of using thin
streams in order to constrain the MW potential.

In performing this study we have obtained the first 6-D phase-space map
for a kinematically cold stellar stream in the Milky Way. We view our present
analysis in same sense as a pilot study for the Gaia \citep{perryman01}
age, when this ESA space mission will deliver dramatically better data on
streams such as GD-1.

This paper is organized as follows: In Section~\ref{sec:stellarpop} we discuss
the analysis of the SDSS photometry, which entails mapping the \GS stream in 3-D
as
well as determining its stellar population properties. In
Section~\ref{sec:kinematics} we present the kinematics, with proper motions
from SDSS-USNOB1.0 and line-of-sight velocities from SDSS and Calar Alto.
In Section~\ref{sec:modeling_all} we combine this information in an iterative
step that involves improved stream membership probabilities, which in turn
affects the estimates of proper motions and distances. This procedure results
in the most comprehensive 6-D data set for a stellar stream in our Milky Way. In
Section~\ref{sec:orbitfitting} we model the stream data by a simple orbit in a
simple parametrized gravitational potential. We measure the potential
circular velocity and find that the overall Milky Way
potential at the \GS stream position is somewhat flattened, but that much of
that flattening can be attributed to the disk.

\section{Stellar population of the stream}
\label{sec:stellarpop}
The probability that a star is a member of the \GS
stream depends on its 6-D
position and its metallicity. In the space of photometric observables, this
means that it depends on $(\alpha, \delta)$, magnitude and color. In practice,
the determination of the stream's angular position, distance and
metallicity
(presuming it is 'old') is an iterative process which we detail here. 

\begin{figure}
	\includegraphics[width=0.5\textwidth]{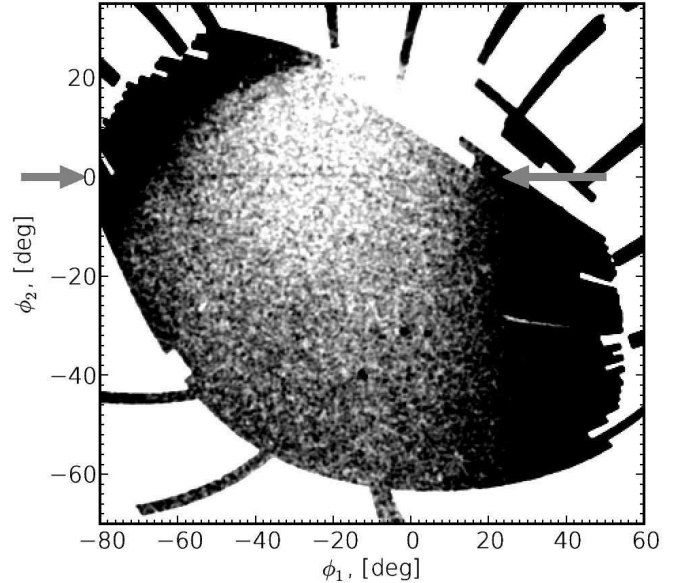}
\caption{The number density of SDSS DR7 stars with $0.15<g-r<0.41$ and $18.1<
r<19.85$, shown in the rotated spherical coordinate system that is approximately
aligned with the \GS stream. The map was convolved with a circular Gaussian with
$\sigma=0.2\degr$. The gray arrows point to the stream, which is barely
visible in this representation, extending horizontally near $\phitwo=0\degr$,
between
$\phione=-60\degr$ and $0\degr$.}
\label{fig:sdss_map_rotated}
\end{figure}

\citet{grillmair06_gd} made the initial map of the stream using a matched
color-magnitude filter based on the CMD of M13 observed in the same filters.
Not presuming a particular metallicity (e.g. that of M13), we start our
analysis with a simple color-magnitude
box selection for stars ($0.15<g-r<0.41$ and $18.1<r<19.85$). The resulting
distribution is shown in Fig.~\ref{fig:sdss_map_rotated}. That particular
color-magnitude box was selected as appropriate to find metal-poor main sequence
(MS)
stars at a distance of $\sim$ 10~kpc, and indeed the stream is marginally
discernible in the Figure. With just a color-magnitude box, however, the
detection fidelity of that stream is
noticeably lower than that achieved by \citet{grillmair06_gd}(their Fig.~1).
The distribution of stars on Figure~\ref{fig:sdss_map_rotated} is plotted in
a rotated spherical coordinate system ($\phione, \phitwo$), approximately
aligned with the stream, where $\phione$ is longitude and $\phitwo$ is the
latitude. The north pole of that
coordinate system is located at $\alpha_p$=34$\degr$.5987, $\delta_p$ =
29$\degr$.7331, the zero-point for $\phione$ is located at $\alpha=200\degr$,
and we will use this coordinate system for convenience throughout
the paper to describe stream positions (the transformation matrix from
($\alpha,\delta$) to ($\phione,\phitwo$) is given in the appendix).

\begin{figure}
  \includegraphics[width=0.5\textwidth]{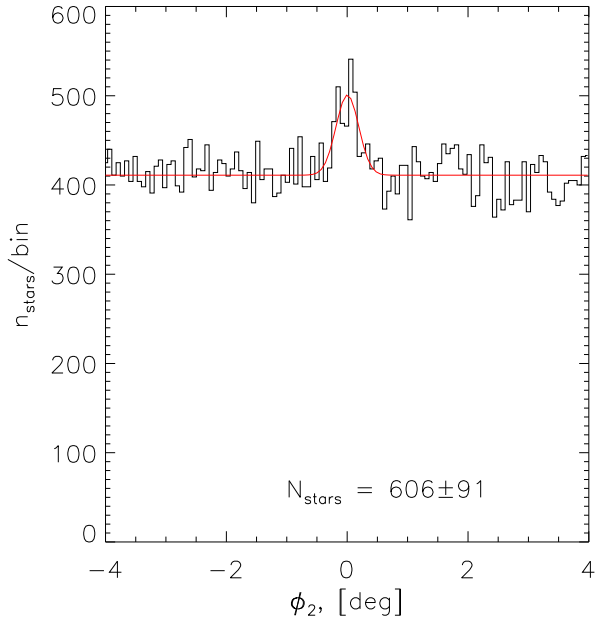}
\caption{One-dimensional stellar density profile across the stream using the
stars with $0.15<g-r<0.41$ $18.1< r<19.85$ across the $\phitwo=0\degr$ axis,
integrated along the stream in the interval $-60\degr<\phione<-10\degr$. The
Gaussian fit with $\sim$ 600 stars and $\sigma$=12\arcmin\ is shown in red.}
\label{fig:stream_profile}
\end{figure}

If we integrate the low-contrast 2D map in Fig.~\ref{fig:sdss_map_rotated}
along the $\phione$ axis, creating a one-dimensional profile of the stream,
the presence of the stream becomes very clear. Figure~\ref{fig:stream_profile}
shows this profile
for stars with $0.15<g-r<0.41$, $18.1<r<19.85$ and with
$-60\degr<\phione<-10\degr$. In that Figure we also overplot the Gaussian fit to
this profile with $\sim$ 600 stars and Gaussian width ($\sigma_{\phitwo}$) of
$\sim$
12\arcmin. This number of stars corresponds to a total stellar mass
of $M_*\approx 2\times 10^4M_\odot$, if we assume a distance of $\sim$
10~kpc (see below), and a Chabrier IMF \citep{chabrier01} with an old,
metal-poor
stellar population. Given that number of stars, we expect to see around 3000
stream stars in SDSS with $r<$22. \change{The mean surface brightness of the
stream is around 29 mag/sq.arcsec.}

\begin{figure*}
\includegraphics[width=\textwidth]{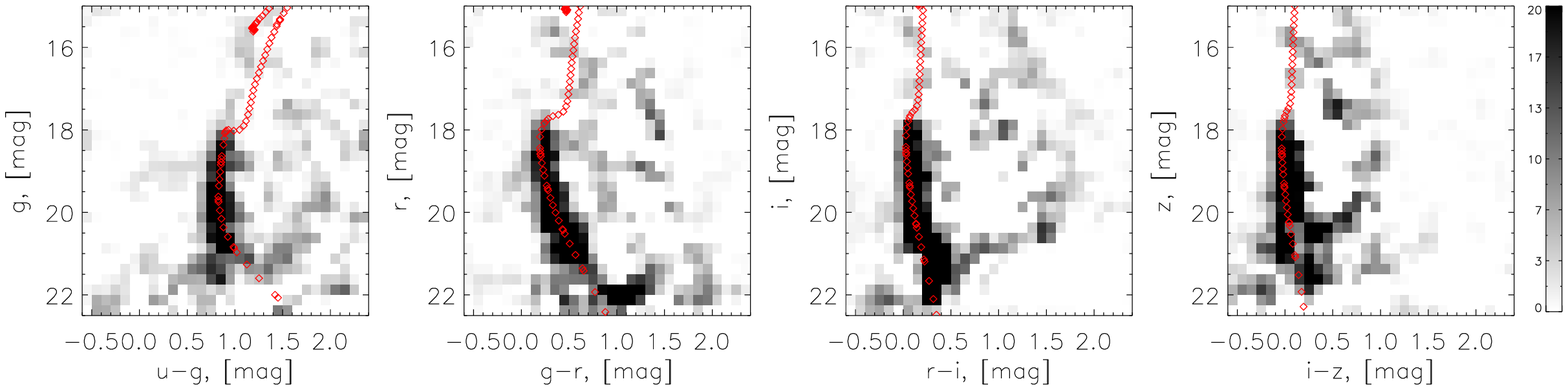}
\caption{Color magnitude (or Hess) diagrams of the stream derived
by statistical background subtraction using the 
Eq.~\ref{eq:stream_gauss} fit, in different filters ($u-g$ vs $g$,
$g-r$ vs $r$, $r-i$ vs $i$ and $i-z$ vs $z$ (from left to right). The grayscale
shows the number of stars per rectangular bins. All the magnitudes are
extinction corrected. Overplotted are theoretical isochrones for
${\rm age}=9$~Gyrs, ${\rm log(Z/Z}_\odot)=-1.4$, ${\rm distance} =8.5$~kpc.}
\label{fig:ugriz_cmds}
\end{figure*}

\begin{figure}
\includegraphics[width=0.5 \textwidth]{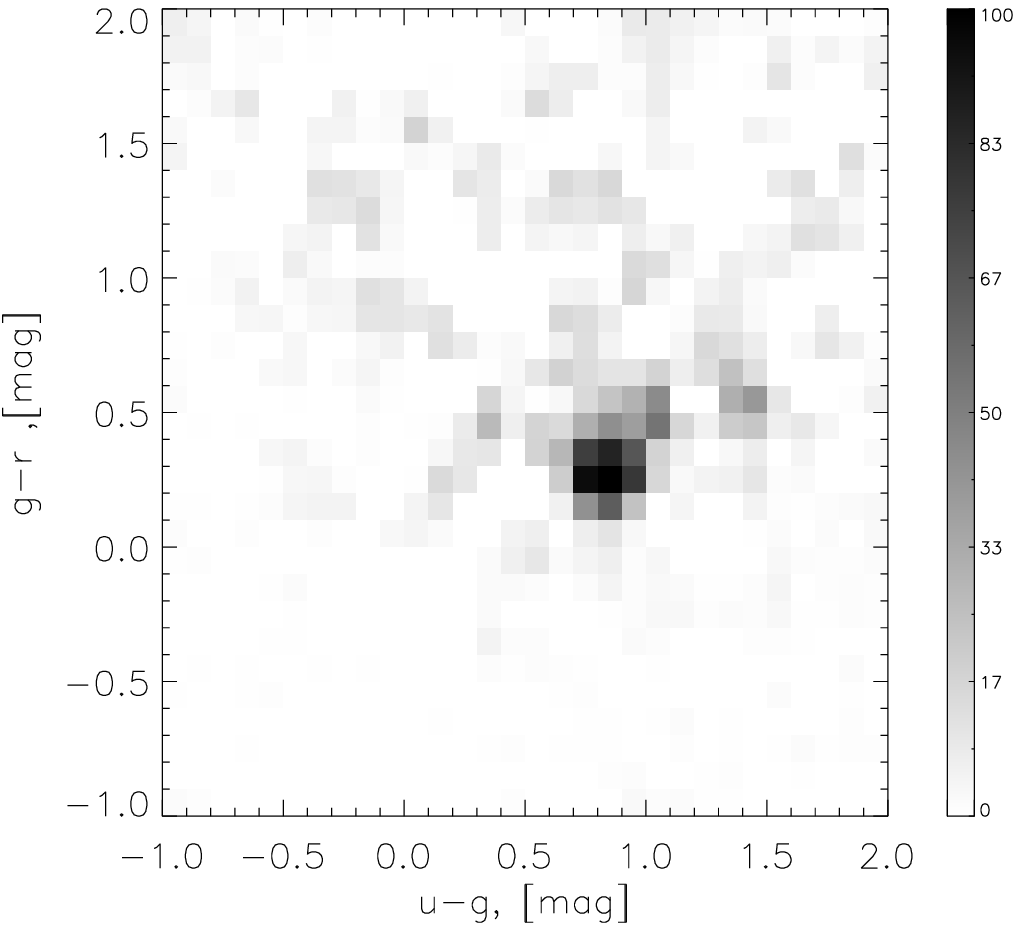}
\caption{$(u-g)-(g-r)$ color-color diagram of the stream, which constitutes a
photometric metallicity estimator \citep[following][Eq.~4]{ivezic08}, shown
after statistical background subtraction as for Fig.~\ref{fig:ugriz_cmds}. All
the magnitudes were extinction corrected. The grayscale shows the number of
stars per bin, with a distinct concentration of stars at (0.8,0.35), that
implies a well defined metallicity $[{\rm Fe/H}]=-1.9\pm0.1$.}
\label{fig:colcol}
\end{figure}

We expand this approach to the determination of the CMD of the stream. The
data and the fit shown in Figure~\ref{fig:stream_profile} was
obtained for a fairly wide color-magnitude selection box. But we can
construct such a profile for any other (e.g.) color magnitude box and that
profile can then be fitted by 
\begin{eqnarray}
N_{obs}(\phitwo | CMD)=N_{BG}(CMD)+\nonumber\\
N_{stream}(CMD)\times\frac{1}{\sqrt{2\pi}\sigma_{\phi 2}}
{\rm exp}\left(-0.5\left(\frac{\phitwo-\phi_{2,0}}{\sigma_{\phi
2}}\right)^2\right)
\label{eq:stream_gauss}
\end{eqnarray} , where
CMD refers to a given color-magnitude bin, and where we assume that both
 center ($\phi_{2,0}$) and width ($\sigma_{\phi 2}$) of the stream are fixed at
0 and 12 arcminutes. A fit of the Eq.~\ref{eq:stream_gauss} model to the
observed data $N_{obs}(\phitwo|CMD)$ can be performed in
$\chi^2$ sense. As a result $N_{stream}(CMD)$,
the number
of stream stars (and its error), can be determined for each given
color-magnitude bin, resulting in a Hess diagram for \change{different pairs of SDSS
filters ($u-g$, $g-r$,$r-i$, $i-z$).}
Figure~\ref{fig:ugriz_cmds} shows the resulting Hess diagram of the stream
derived in several bands. These clearly show a main
sequence(MS). The location of the MS turn-off cannot be clearly identified,
although there may be a hint at $g=18.5$, $u-g=1$. On
Figure~\ref{fig:ugriz_cmds} we also overplot the \citet{marigo08} isochrones
with age$=9$~Gyrs, ${\rm log(Z/Z}_\sun)=-1.4$ at 8.5~kpc, which seem to match
quite well.
\citet{ivezic08} recently showed that the location of MS stars in the
$(u-g)-(g-r)$ color-color plane is a good metallicity diagnostic. Therefore, we 
construct the $(u-g)-(g-r)$ color-color
diagram of the stream stellar population shown in Fig.~\ref{fig:colcol},
which exhibits a distinct concentration of stars at (0.8,0.3). This argues
for a population of single or a dominant metallicity and we can convert this
color location to a metallicity using Eq.~4 from~\citet{ivezic08}:
${\rm [Fe/H]}_{phot}=-1.9\pm 0.1$. This provides a metallicity estimate that
is directly linked to SDSS spectral metallicity estimate.

\begin{figure*}
\includegraphics[width=\textwidth]{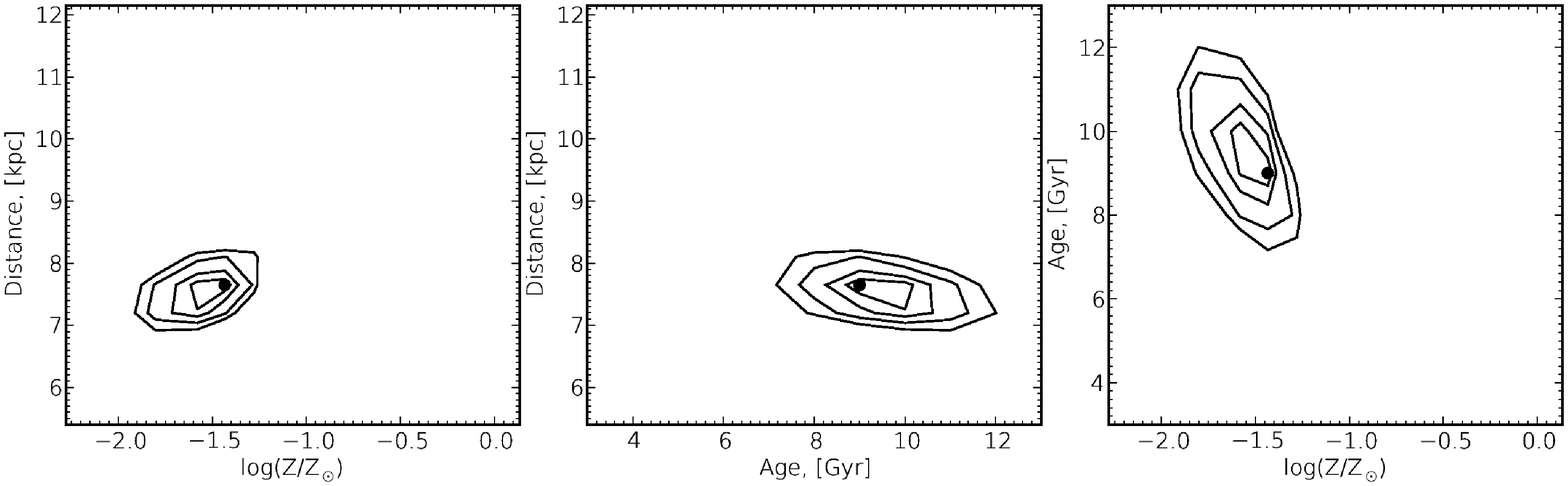}
\caption{Isochrone fits to the color magnitude
diagrams (Fig.~\ref{fig:cmd_split}) in $u$, $g$, $r$, $i$, $z$ bands for the
stream integrated over $-60\degr<\phione<-10\degr$. The contours show the formal
60\%, 90\%, 99\%, 99.9\% confidence regions as a function of distance, age and
metallicity respectively. Filled circles show the location of the best goodness
of fit point.}
\label{fig:cmd_fit}
\end{figure*}

\begin{figure}

\includegraphics[width=0.5 \textwidth]{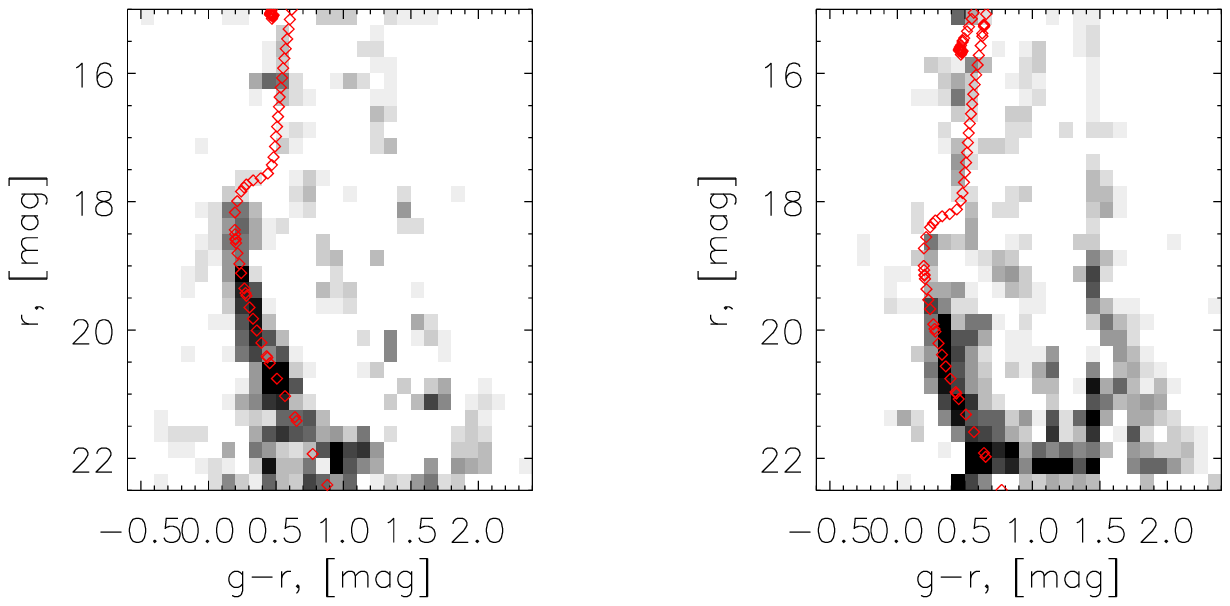}
\caption{Distance variation along the stream. The CMD diagrams of the stream for
two different parts of the stream, $-40\degr<\phione<-20\degr$ (left),
$-10\degr<\phione<10\degr$ (right). The isochrones for the the best fit model
${\rm log(Z/Z}_\odot)=-1.4$, age$=9$~Gyr were shifted to the distance of 8.5~kpc
on the left panel and
to 11~kpc on the right panel. Some distance variation apparent, with the stellar
population shown on the right located at greater distances.}
\label{fig:cmd_split}
\end{figure}

To derive the metallicity, age, and distance of stream stars in a systematic
way, we fit the color-magnitude diagrams  using a
grid of isochrones populated realistically according to the IMF
\citep{dolphin02,dejong08}. We focus on fits to the
color-magnitude diagrams in $u$, $g$ and $g$, $r$ filters, since that the
$u-g$ color of the MS turn-off is a good metallicity indicator
\citep{ivezic08}. We create the synthetic Hess
diagrams for a grid of model stellar
populations \citep{girardi00,marigo08}\footnote{To retrieve the isochrones we
used the web interface provided by Leo
Girardi at the
Astronomical Observatory of Padua
\url{http://stev.oapd.inaf.it/cgi-bin/cmd\_2.1}} with different ages
($3-12$~Gyr), metallicities (${\rm Z}=0.0001-0.025$), distances ($6-14$~kpc),
and a Chabrier IMF\citep{chabrier01}.
We then explore that grid by computing log-likelihood of the distribution of
stars in color-magnitude space. Figure~\ref{fig:cmd_fit} shows the 2D profile
likelihoods contours 
of the age vs metallicity, age vs distance and distance vs metallicity planes.
The filled circle indicates the best fit model: age$=9$~Gyrs,
${\rm log(Z/Z}_\sun)=-1.4$ and distance$=8$~kpc.
Clearly the age is the least well constrained parameter; the distance
seems to be relatively well constrained, but has a covariance with [Fe/H]. We
will revisit this issue later, as the analysis of Fig.~\ref{fig:colcol} implies
a lower metallicity. Fig.~\ref{fig:ugriz_cmds} shows that the isochrones are
reproducing the observed
Hess diagrams well, and hence further in the paper we will use  t$=9$~Gyrs,
${\rm log(Z/Z}_\sun)=-1.4$ as the baseline model for the stream's stellar
population. It should be noted that the distance
measurement from Fig.~\ref{fig:cmd_fit} represents the averaged
distance along the stream from $-50\degr<\phione<-20\degr$. In
section~\ref{sec:modeling_all}  we will present estimates of the
distance to different parts of the stream. 

It is noticeable that the metallicity derived from the CMD fitting is higher
than from the estimate based on empirical calibration of \citet{ivezic08}(see
above) and higher than the measurement based on the SDSS spectra given by
\citet{willett09}. This discrepancy is understandable given the known
inaccuracies
of the isochrones in the SDSS photometric system \citep{an08}. In
particular the Figure~19 of \citet{an08} paper clearly shows a mismatch between
fiducial isochrone derived for the M92 globular cluster (which is used elsewhere
as a good approximation of old metal-poor stars in the halo) and the theoretical
isochrones. \change{For main sequence stars below the turn-off (which are of
the most interest here), the mismatch between a fiducial isochrone of the
M53 globular cluster (which has metallicity [Fe/H]$\sim -2$) from \citet{an08} and the
isochrone, which we are using can be approximated by a distance shift of
$\sim$ 10\%. Therefore in our analysis we reduce all the distances derived on
the basis of the CMD fit by 10\%. Due the these described inaccuracies 
with the isochrones in SDSS filters, some remaining systematic error in
distances may still exist, although the study of \citet{eyre09_gd1} seems to
indicate it is small. Until isochrones in SDSS filters are fixed, the usage of
fiducial isochrones may give more consistent results, but in the current paper
we decided to proceed with the theoretical isochrones and 10\% distance
correction.}

Splitting the CMD data into $\phione$ bins shows that there is a distance
gradient
along the stream: Figure~\ref{fig:cmd_split} shows two Hess
diagrams obtained for two different pieces of the stream, on the left
$-40\degr<\phione<-20\degr$, and on the right\ -$10\degr<\phione<10\degr$.
The baseline isochrone is
shifted to distances of 8.5~kpc(left) and 11~kpc(right) respectively. Clearly,
the part of the stream at $-10\degr<\phione<10\degr$ is further from the Sun,
\citep[as already noted by][]{grillmair06_gd,willett09}. 

\begin{figure}
 \includegraphics[width=0.5\textwidth]{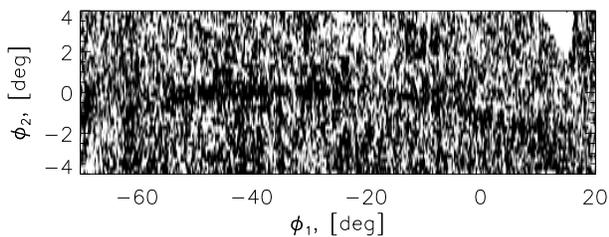}
\caption{Map of stream stars in the rotated coordinate system after
applying the matched CMD filter from Section~\ref{sec:stellarpop}. The figure
shows the histogram of stars in $\phione, \phitwo$, where each star has been
weighted by the CMD weight
$\frac{P_{stream}(u-g,g-r,r-i,r)}{P_{BG}(u-g,g-r,r-i,r)}$. \change{The linear
variation of distance was assumed}}.
\label{fig:stream_match_filter}
\end{figure}

The determination of the CMD properties for the stream allows us to select
the possible stream member stars
with much less background contamination, compared to a simple color
magnitude box. Figure~\ref{fig:stream_match_filter} shows the map of
the stream
(in $\phione,\phitwo$ coordinates) after applying a matched filter based on
the CMD of the stream. For Fig.~\ref{fig:stream_match_filter} each star in
the SDSS dataset was weighted by
the ratio of the stream membership probability and the background probability
$\frac{P_{stream}(u-g,g-r,r-i,r)}{P_{BG}(u-g,g-r,r-i,r)}$ 
,where $P_{stream}(u-g,g-r,r-i,r)$ have been computed based on the stellar
population fit (Fig.~\ref{fig:cmd_fit}), and $P_{BG}(u-g,g-r,r-i,r)$ was
computed empirically from the regions adjacent to
the stream \citep[see e.g.][for the application of similar method]{rockosi02}.
The resulting image after the CMD weighting shows the stream with 
obviously greater contrast than Figure~\ref{fig:sdss_map_rotated}.
\change{It is noticeable that the density of stars in the stream does not
seem do be constant and even does not change monotonically along the stream,
instead the stream seems to have a few clumps and holes. The nature of the
substructure
in the stream is unclear. It may be related to either the history of the
disruption process \citep{kupper10}, or e.g. interaction with dark matter
subhalos around MW \citep{carlberg09}.}

\section{Kinematics}
\label{sec:kinematics}
In this section we describe how we derived estimates of the 3-D kinematics of
the
stream by looking at the proper motions and  radial
velocities of the probable stream members.

\subsection{Proper motion}

Despite the distance of $\sim$ 10~kpc to the \GS we demonstrate in this section
that it is possible to derive good constraints on its proper motion.
The analysis is based on proper motions
derived from combining  USNO-B1.0 \citep{monet03} with SDSS data 
\citep{munn04,munn08}, which we take from SDSS DR7\citep{abazajian09}.

\change{At a distance of $\sim$ 10~kpc a stream that is moving perpendicular to
the line sight with the velocity of 200~km/s (roughly a characteristic velocity
in the halo) should have a proper motion of the
order of $\sim 4$~mas/yr.} Hence the
expected proper motion is
comparable to the precision of individual proper motion measurements
\change{ of $3-4$~mas/yr \citep{munn04}}.
As stream member stars that are adjacent on the sky should have (nearly)
identical proper motions, we can, however, determine statistical proper
motions for ensembles of stars.

\begin{figure*}
\includegraphics[height=0.43\textwidth]{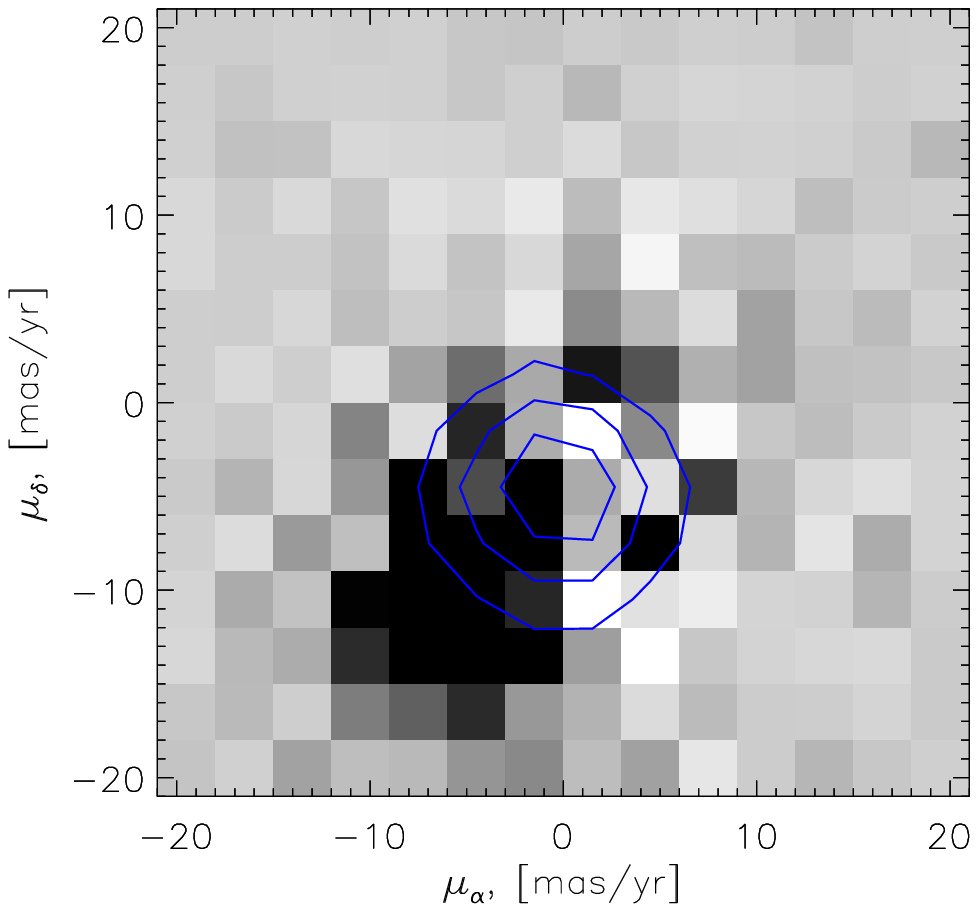}
\includegraphics[height=0.43\textwidth]{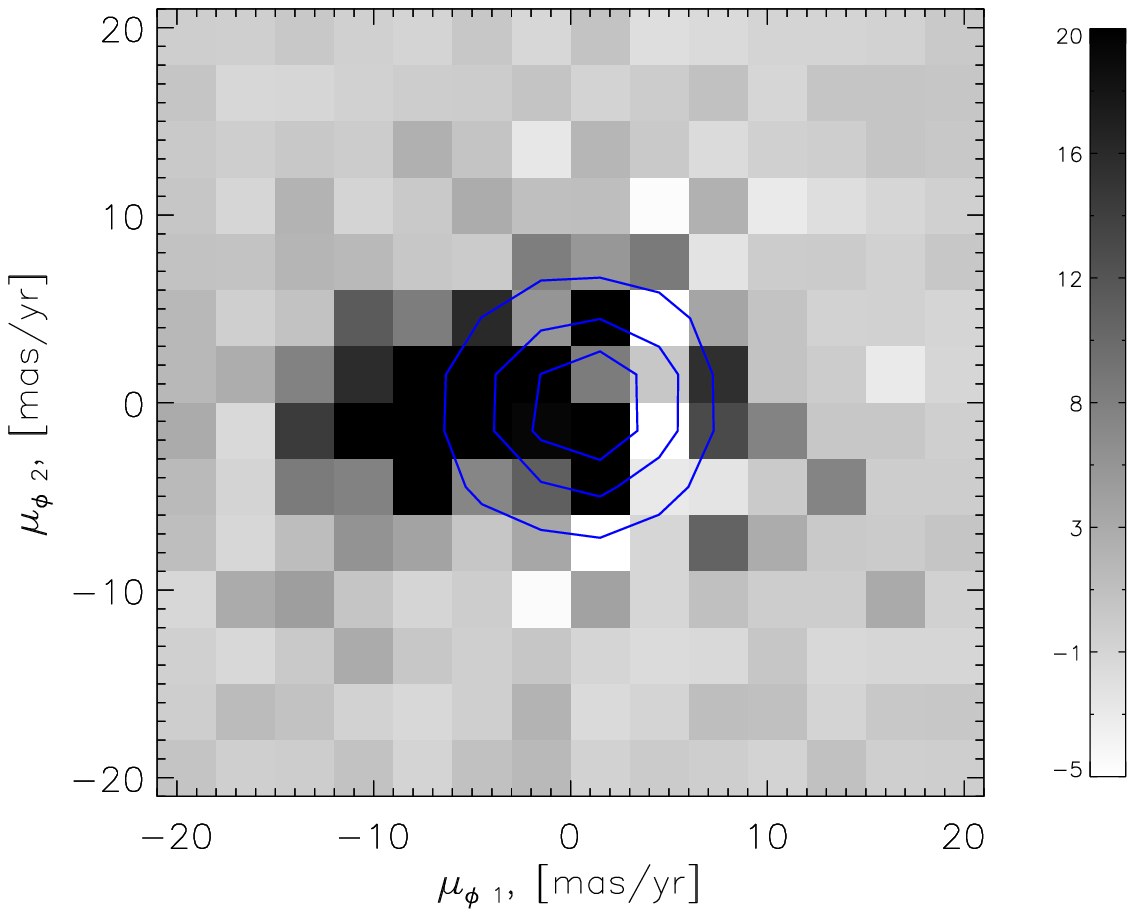}
\caption{Proper motion of the stream. The left panel shows the proper motion in
right ascension and declination $\mu_\alpha, \mu_\delta$ (as observed, e.g. no
correction for the Solar motion in the Galaxy was made). The right panel shows
the proper motion in the rotated coordinate system $(\phione, \phitwo)$
($\phione$
is oriented along the stream) and after the subtraction of the proper motion due
to the Sun's motion in the Galaxy (assuming for now $V_c=220$~km/s). The
grayscale in each of the panels shows the number of stars per
proper motion bin.
\change{ Contours corresponding to 30,60,90 stars per bin show the the
proper motion distribution for the field stars.}
with similar colors and magnitudes to the stream stars. The proper motions of
the stream stars are clearly
distinguishable from the proper motions of the background stars. The right panel
confirms the fact that the stream stars are moving approximately along its
orbit ($\mutwo\approx0$~mas/yr), while the mean proper motions of background
stars after subtracting Sun's
motion are consistent with zero.}
\label{fig:pms}
\end{figure*}

We start by deriving the $\vec{\mu}$-distribution of likely member stars, by
extending the analysis of Section~\ref{sec:stellarpop} and using both the
angular position on the sky (specifically $\phitwo$) and the location in CMDs
for each star to evaluate its stream membership probability.
\change{Specifically, we grid the proper motion space into 'pixels', then we
select the stars within each proper motion 'pixel' and with high membership
probabilities in the CMD ($u-g$,$g-r$,$r-i$,$i$) space $P_{stream}/P_{BG}
\gtrsim 0.1$.
For that sample of stars within each proper motion 'pixel' $(\vec{\mu},
\vec{\delta \mu})$,  we determine $N_{stream}(\vec{\mu}|\phitwo, CMD)$ (number
of stream stars) by integrating along the stream direction $\phione$ and fitting
the resulting $\phitwo$ distribution with the Eq.~\ref{eq:stream_gauss}.}
The resulting
distribution $N_{obs}(\vec{\mu})$ is shown in Fig.~\ref{fig:pms}. The grayscale
in the left panel of the Figure shows the proper motion distribution of 
$\mu_\alpha, \mu_\delta$ of probable stream member stars ($N_{stream}$)
integrated along
the stream, while the contours show the proper motion distribution of the
background stars selected with the same color-magnitude
criteria (corresponding to $N_{BG}$ from Eq.~\ref{eq:stream_gauss}). It is clear
that the stream stars
are on average moving differently than the bulk of background stars. However,
the observed proper motions contain the reflex motion of Sun's motion in the
Galaxy. We can account for this and then convert $\vec{\mu}$
to $(\muone,\mutwo)$, the proper
motion along the stream $\muone\equiv\mu_{along}$ and proper motion across
the stream $\mutwo$ in the Galactic rest-frame (where $\phione, \phitwo$ are
the stream coordinates introduced in Section~\ref{sec:stellarpop}). The proper
motion component arising from the Sun's movement in the Galaxy can be easily
computed for each star. 
$$\vec{\mu}_{reflex}=\frac{1}{4.74\, \lvert\vec{r}\rvert}
(\vec{V}_\odot-(\vec{V}_\odot\cdot\vec{r})\frac{\vec{r}}{ \lvert\vec { r }\rvert
^2 } )$$
where $\vec{V}$ is a 3-D velocity of the sun ($\sim$ 220~km/s) and $\vec{r}$ is
the
vector from the sun towards each star. As we approximately know the distance
to the stream, this correction
$\mu_{\phi_{1,2},c}=\mu_{\alpha,\delta}-\mu_{reflex}$ can be done. We will
discuss the consequences of the uncertainties in the Sun's motion and the
stream differences in Section~\ref{sec:modeling_all}. 

The right panel of Fig.~\ref{fig:pms} shows the distribution of
$\muone,\mutwo$ of the  stream stars. The contours show the corresponding
distribution of background stars with the similar color-magnitudes to the
stream stars. Reassuringly we see that stream stars are moving approximately
along
$\phione$, i.e. along the stream orbit, an important plausibility check for the
correctness of the proper motion measurement. In contrast, the
proper motion distribution of the background stars after subtracting the proper
motion due to sun's movement is centered around $(\muone,\mutwo)=(0,0)$, which
appears reasonable since with our color-magnitude selection we are selecting
primarily
the halo stars at distances $\sim 10$~kpc. Those show little net
rotation\citep{carollo07,xue08}. The estimate $\langle\muone\rangle\approx
-8$~mas/yr also immediately implies that the stream is going retrograde
with respect to the Milky Way's  disk rotation.

\begin{figure}
\includegraphics[width=0.5\textwidth]{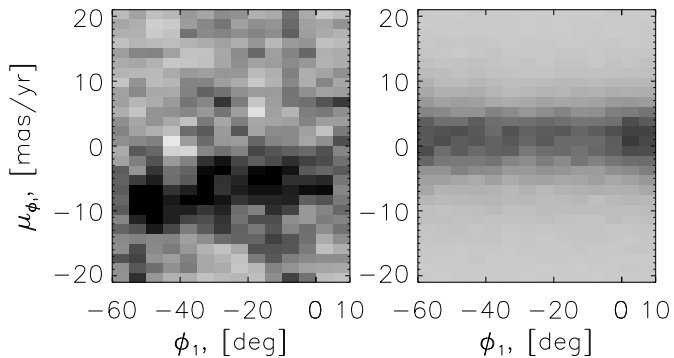}
\caption{Variation of the proper motions along the stream (corrected for the
Solar reflex motion, assuming $V_0=220$~km/s). The left
panel shows the distribution of $\muone$ for the stream candidate stars as a
function of
$\phione$. The right panel shows the distribution of $\muone$ of the field stars
selected using the same color-magnitude criteria as the stream stars. The
variation of proper motions of stream stars with $\phione$ is clearly
visible in the left panel. Near $\phione\sim 0\degr$ the proper motion of the
stream is around
$-5$~mas/yr, while at $\phione\sim-50\degr$ it is around $-8$~mas/yr
\change{(corresponds to $\sim 300$~km/s at $8$~kpc)}.}
\label{fig:pm_var}
\end{figure}

In Figure~\ref{fig:pm_var} we illustrate the
proper motion  variation along the stream. These plots, showing only
$\muone(\phione)$ were derived the same way as Fig.~\ref{fig:pms}, except that
we did not integrate in $\phione$ along the entire stream but only in
$\phione$ intervals. The right panel of Fig.~\ref{fig:pm_var} shows the
distribution of the proper motions along the stream of the background stars.
The left panel of the Figure shows the distribution of proper motions of
likely stream member stars as a function of angle along the stream (the proper
motion due to the Sun's motion was subtracted). The left panel reveals a
slight, but significant gradient in $\langle\muone\rangle$, of the order of
3 to 5~mas/yr.
Note that the decrease of the
proper motions towards $\phione = 0$ coincides with the distance
increase to the stream (see Fig.~\ref{fig:cmd_split}).

Having determined the stream proper motions, we can further improve the
CMD-filtered map of the \GS stream (Fig.~\ref{fig:stream_match_filter}) by
requiring that the proper motions of the stars are consistent with the proper
motion of the stream. That is shown in
Fig.~\ref{fig:stream_map_pm}. and discussed in Section~\ref{sec:velocities}.

\subsection{Radial velocities}
\label{sec:velocities}
To construct the 6-D phase space distribution of the stream, the radial
velocities are the remaining datum. By necessity the actual sample for which
spectra, and hence radial velocities, will be available, will differ from the
photometric sample just described. In this section we will use
both the data from the SDSS/SEGUE survey \citep{yanny09} as well as radial
velocities obtained by us with the TWIN spectrograph on Calar Alto,
specifically targeting likely stream member stars. 

\subsubsection{SDSS radial velocities}

\begin{figure*}
\includegraphics[width=1\textwidth]{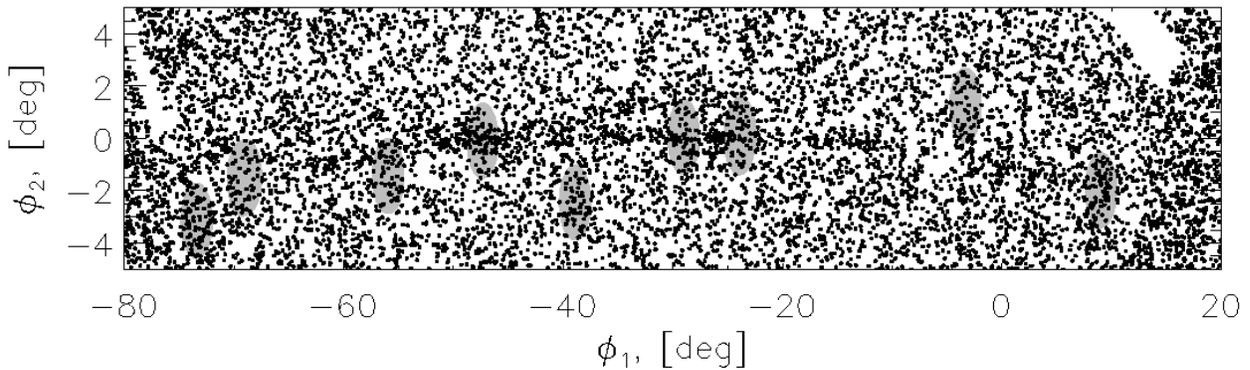}
\caption{Stars that match the expectations for stream members with regards
to proper motions \change{($-15$~mas/yr$<\muone<-5$~mas/yr,
$|\mutwo|<3$~mas/yr)}, colors and magnitudes (but no $\phitwo$ filter), used in
the
candidate selection for radial velocity measurements. The
stream can be clearly seen in distribution of individual stars. The locations of
the SEGUE DR7 pointings are shown by gray circles.}
\label{fig:stream_map_pm}
\end{figure*}

SEGUE and SDSS only provide sparse spatial sampling of high latitude
stars. SEGUE did not target any \GS member stars specifically. Therefore we
have
to search through the existing SEGUE spectra to identify likely, or
possible, member stars by position on the sky, CMD position and proper motion.
In the previous section we described that we used the ratio of the
stream/background
probabilities $\frac{P_{stream}(u-g,g-r,r-i,r)}{P_{BG}(u-g,g-r,r-i,r)}$
to select high probability members of the stream. Now additionally to that we 
also select the stars within the $\mu_\alpha,\mu_\delta$ box (see
Fig.~\ref{fig:pms}). That allows us to have a sample of stream
stars with much less background contamination, although the overall size of
that sample is significantly smaller, since the the SDSS/USNO-B1.0 measurements
 of the proper motions were done for stars with $r\lesssim
20$\citep{munn04}. To illustrate how good the proper motion selection is when
combined
with the color-magnitude  selection we show map of high probability
stream member stars on Figure~\ref{fig:stream_map_pm}. The stream is now clearly
seen in
individual stars. In Figure~\ref{fig:stream_map_pm} we
also overplot the location of existing SEGUE DR7 pointings,
some of which  cover the stream. Therefore we may expect to find some stream
members among the SEGUE targets
in these fields.

\begin{figure}
\includegraphics[width=0.5\textwidth]{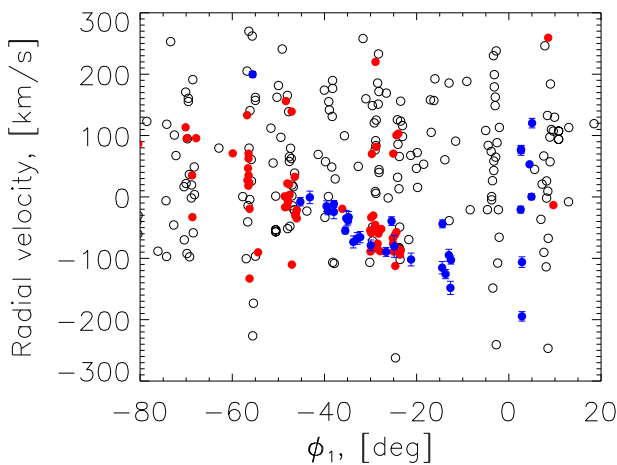}
\caption{Radial velocities of likely stream stars (filled circles). The
Figure shows the radial velocities drawn from the SDSS/SEGUE (red circles) and
Calar Alto spectra (blue circles). The red and blue symbols reflect the radial
velocities of all stars matching in color, magnitude and proper motion that
have positions with $|\phitwo|<0.3\degr$. The open circles represent
SDSS/SEGUE velocities of similar stars but with $|\phitwo|>0.3\degr$. The
radial component of the Solar reflex motion (taking $V_0=220$~km/s) was
subtracted from all datapoints.}
\label{fig:radvels} 
\end{figure}

Figure~\ref{fig:radvels} shows the SDSS/SEGUE radial velocity distribution as a
function of $\phione$ for those stars whose proper motions and color-magnitude
position are consistent with stream membership, and which are located
within 3 degrees from the center of the stream. \change{The typical uncertainty 
of the SDSS/SEGUE radial velocities is $\sim$20~km/s.} The
filled red circles
in this Figure show the subset of stars located within 0$\degr$.3 from the
center of the stream and therefore represent the subset with very high
membership probability, while the open black circles represent(spatially
selected, $|\phitwo|>0.3\degr$) background stars with similar proper motion and
color magnitude.
The filled symbols in Fig.~\ref{fig:radvels} clearly delineate the radial
velocity of the stream. Clumps of red circles are visible at
($\phione,V_{rad}$)$\approx$(-25\degr,-100~km/s), (-30\degr,-80~km/s),
(-47\degr,0~km/s), (-55\degr,40~km/s). 
In order to perform the formal measurements of the radial velocities we
performed a maximum likelihood fit by a model, consisting from two Gaussians
one (wide) Gaussian was representing the background distribution, while the
second (narrow) was modeling velocity distribution of stream stars. This fit
gave us the following results: 
($\phione,V_{rad}$)=(-56\degr,39$\pm$14~km/s), (-47\degr,-7$\pm 10$~km/s),
(-28\degr,-61$\pm$6~km/s), (-24\degr, -83$\pm$9~km/s).

\subsubsection{Calar Alto radial velocities}

Since the SDSS/SEGUE radial velocities only provide constraints at a few
points of the stream, we decided to obtain additional radial velocity
information with targeted observations. We based the target selection for
likely stream members on all the available information discussed in the
previous sections: position on the sky, color-magnitude location and proper
motions. We selected 34 stars likely members with $r \lesssim$ 19
for the observations. All these stars are within the sample plotted in the
Figure~\ref{fig:stream_map_pm}. \change{These stars are mostly main-sequence and
main-sequence turn-off stars with $18\lesssim r \lesssim 19$}.

The observations were performed the TWIN spectrograph on the 3.5m telescope at
Calar Alto observatory, during several nights of service observing in
February 2009. \change{The TWIN spectrograph is the intermediate resolution
spectrograph installed in the Cassegrain focus of the telescope. It consists
from two separate spectroscopic channels (blue and red) behind the common
entrance slit. The light
from the slit is splitted into ``red'' and ``blue'' beams by a dichroic mirror.}
We used the blue and red arms of the
spectrograph at a resolution of $4000-5000$ to observe the H$_\beta$, Mgb lines
and CaII near-IR triplet respectively. The standard data reduction steps were
applied to the dataset using custom written routines in Python language.
\change{The median signal to noise ratio per pixel was 7 for the blue spectra
and 3 for red spectra.}

We used both the blue and the red spectra to compute the radial velocity of
each star. The radial velocity of each star was derived by minimizing 
$\chi^2$  as a function of velocity shift of the template convolved
with the appropriate Line Spread Function (LSF). The $\chi^2$
for each star was a sum of the $\chi^2$ for the blue and the red part of
the spectra. As template in the blue spectral range we used  the  spectra from
the ELODIE database \citep{prugniel07} for stars
of similar color and magnitude to the targeted ones and with low metallicity
$[{\rm Fe}/{\rm H}]\sim -2$. In the red spectral range the template spectra was
simply
consisting from three lines of Ca triplet at 8498.02\AA, 8542.09\AA\ and
8662.14\AA. The error of each
velocity measurement was determined using the condition
$\Delta(\chi^2(V))=1$.

The Calar Alto measurements of the velocities together with their errors are
overplotted in blue symbols in Fig.~\ref{fig:radvels}. It is apparent from
Fig.~\ref{fig:radvels} that for $-50\degr<\phione<-10\degr$ the targeting
strategy
was very successful, nicely delineating the projected velocity gradient along
the stream. Overall out of 34 observed stars $\sim$ 24
stars belong to the stream and $\sim$ 5 didn't have enough S/N for the
velocity determination. Unfortunately the targeting near $\phione \approx
+5\degr$ failed
to identify stream members, probably because the stream there is less intense
and further away. 

\section{Modeling}
\label{sec:modeling_all}
In the previous section we described the
derivation of different stream properties such as distance,
position on the sky, proper motion separately. In this Section we will map the
stream in 6-D position-velocity space in a more consistent way, using all the
available information \citep[see e.g.][for the application of similar, although
simpler method to Sgr stream]{cole08}. This will provide us with a set of orbit
constraints along different sections of the \GS stream, which we will then model
by an orbit to derive potential constraints. 

\subsection{Positions on the sky and distances to the stream}

We start by characterizing the projected stream position and its distance
from the Sun through a maximum likelihood estimate for a parametrized model of
the
stream $P_{stream}(r,g-r,\phione,\phitwo)$ that describes it in 4-dimensional
space of photometric observables
$r$,$g-r$,$\phione$,$\phitwo$:

\begin{eqnarray}
P_{stream}(r,g-r,\phione,\phitwo)  = 
{\rm CMD}(r,g-r,D(\phione)) \nonumber \times \\
\times I(\phione) \times 
\frac{1}{\sqrt{2\,\pi}\sigma_{\phitwo}(\phione)} {\rm
exp}\left(-\frac{(\phitwo-\phi_{2,0}(\phione))^2}{2\,\sigma_{\phitwo}
^2(\phione) } \right) 
\end{eqnarray}
Here $\phi_{2,0}(\phione)$ is the $\phitwo$ position of the stream center on
the sky as a function of $\phione$, $\sigma_{\phitwo}(\phione)$ is the
projected width of the stream in $\phitwo$,
$I(\phione)$ is the
``intensity'' (i.e. the number density) of the stream as a function of
$\phione$, and $D(\phione)$ is the
distance to the stream. Further,
$CMD(r,g-r,D(\phione))$ is the normalized
Hess diagram (i.e. the probability distribution in CMD space) expected for
the stream's stellar population at a
distance of $D(\phione)$ after accounting for the observational errors. We
construct that CMD based on the age and metallicity obtained in
Section~\ref{sec:stellarpop} and the isochrones from~\citet{girardi00,marigo08}
(assuming that $\frac{d[Fe/H]}{d\phione}$=0 and $\frac{d(age)}{d\phione}$=0).
In this model, $P_{stream}$ depends on four
functions -- $I(\phione)$, $\phi_{2,0}(\phione)$, $\sigma_{\phitwo}(\phione)$,
$D(\phione)$ -- which we take to be piecewise constant; i.e. for intervals
$\delta\phione$ they simply become four parameters.

For the field stars, $P_{BG}(r,g-r,\phione,\phitwo)$ the analogous 4D distribution 
is
$$P_{BG}(r,g-r,\phione,\phitwo)= I_{BG}(\phione,\phitwo)\times
CMD_{BG}(r,g-r,\phione)$$,
where $I_{BG}(\phione,\phitwo)$ is the 2D number density distribution of the field stars around
the stream and $CMD_{BG}(r,g-r,\phione)$ the corresponding 
color-magnitude diagram. \change{These functions are determined
empirically from the data in parts of the sky adjacent to the stream
($|\phitwo|\gtrsim0.5\degr$).}
$I_{BG}(\phione,\phitwo)$ is determined by fitting the
density of the stars in the $\phione,\phitwo$ space by a polynomial.
$CMD_{BG}(r,g-r,\phione)$ is determined by constructing the Hess diagrams using
all the stars with $0.3\degr<|\phitwo|<5\degr$ in
several $\phione$ bins.

To simplify the determination of $P_{BG}(r,g-r,\phione,\phitwo)$ and
$P_{stream}(r,g-r,\phione,\phitwo)$, we split the stream in several $\phione$ pieces,
and consider $I(\phione)$, $\sigma_{\phitwo}$, $D(\phione)$ and
$\phi_{2,0}(\phione)$ as constants within them. The log-likelihood for the
mixture of the $P_{stream}$ and $P_{BG}$ distribution can be written as (here
for convenience we introduce $\alpha$ as a fraction of stream stars instead of
$I(\phione)$):

\begin{eqnarray}
{\rm ln}(\mathcal L) = \sum\limits_{stars} 2\,{\rm ln} (\alpha\,
P_{stream}(r_i,g_i-r_i,\phi_{1,i},\phi_{2,i})+ \nonumber \\
(1-\alpha)P_{BG}(r_i,g_i-r_i,\phi_{1,i},\phi_{2,i}))
\end{eqnarray}

and should be maximized with respect to the parameters ($\sigma_{\phitwo}$,
$D(\phione)$, $\phi_{2,0}(\phione)$ and $\alpha$). The maximization is
performed using the Truncated Newton method\citep{nash84}. \change{The 
parameter errors are obtained using numerically computed Fisher
information \citep[see e.g.][]{cox74}.}

The left and central panels of Figure~\ref{fig:obsdata} show the resulting
estimates of the projected position and the distance of the stream, the
parameters used in the subsequent orbit fitting. We  do not use the number
density of stream stars, as it varies noticeably along the stream (see
Fig.~\ref{fig:stream_match_filter}) and the reason of these variations is not
clear. It is apparent from Fig.~\ref{fig:stream_map_pm} that the
projected stream position is very well defined, and that a distance gradient
exists along the stream. 

\subsection{Proper motions}

\begin{figure*}
 \includegraphics[width=\textwidth]{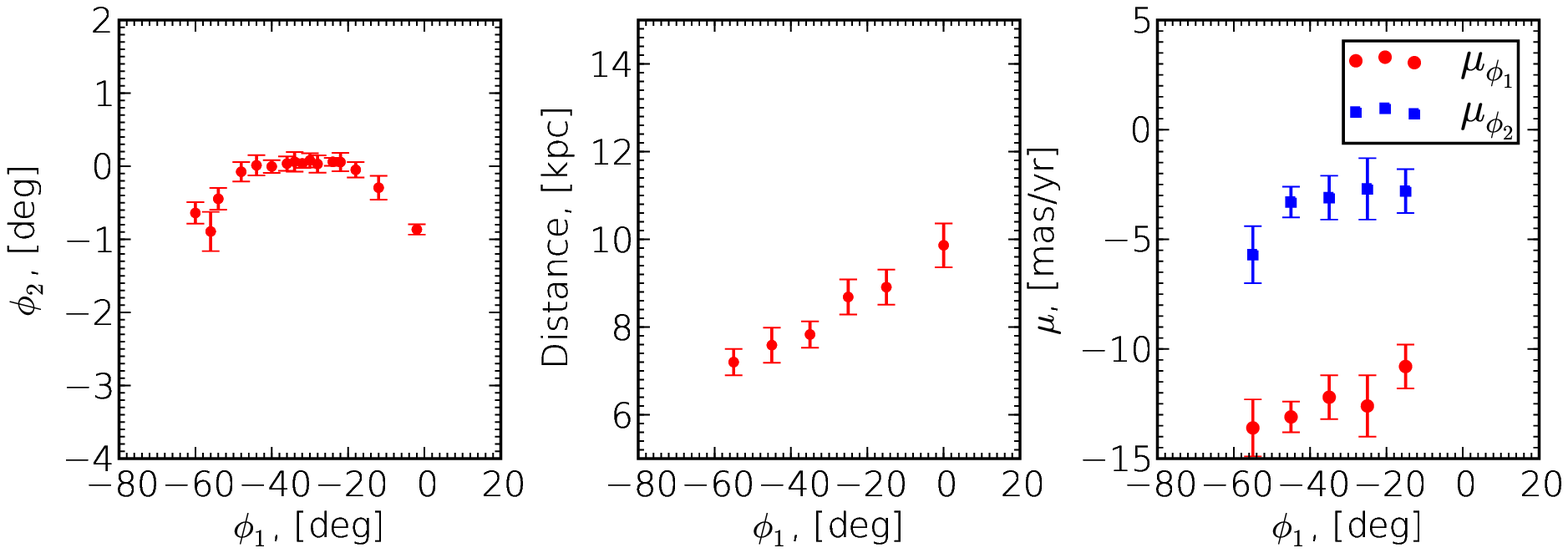}
\caption{Summary of photometrically derived stream properties based on maximum
likelihood
fits to chunks of the stream, drawing on SDSS photometry and astrometry (see
Section~\ref{sec:modeling_all}). The left panel shows the positions of the
stream on the
sky in $\phione$,$\phitwo$ coordinates. The middle panel show the
measurements of the distances as a function of $\phione$.
The right panel
shows the statistical proper motions of the stream stars (without the
correction for the Solar motion); red circles
show the $\mu_{\phione}$ (the proper motion along the stream), blue squares
show the $\mu_{\phitwo}$ (the proper motion perpendicular to the stream).}
\label{fig:obsdata}
\end{figure*}

The likelihood maximization
just described also results in stream membership probabilities for any given star i,
$P_{stream}(r_i,g_i-r_i,\phi_{1,i},\phi_{2,i})$. This information can then be
used to estimate via maximum likelihood the mean proper motions of different stream pieces, thereby extending the observational estimates to the full 6-D space
($r$,$g-r$,$\phione$,$\phitwo$,$\muone$,$\mutwo$). 

\begin{eqnarray}
P_{stream,\mu }(r,g-r,\phi_{1},\phi_{2},\muone,\mutwo) = \nonumber\\
P_{stream}(r,g-r,\phione,\phitwo)  \times 
\frac{1}{2\,\pi\,\sigma^2_\mu}\times \nonumber \\
{\rm exp}\left(-\frac{(\muone-\mu_{\phione,0}(\phione))^2+(\mutwo-\mu_{\phitwo ,
0 } (\phione))^2 } { 2\,\sigma^2_\mu} \right)
\end{eqnarray}

where we simply take the previously determined $P_{stream}$ as a prior, to be
modified by the
Gaussian distribution in the 2D proper motions space. Here, the proper
motion distribution is characterized by three functions,
$\mu_{\phione,0}(\phione)$, $\mu_{\phitwo , 0 }
(\phione)$ and  $\sigma_\mu$, which again we take to be piecewise constant. The
distribution of the background stars in the 6-D space
$P_{BG,\mu}(r,g-r,\phione,\phitwo,\muone,\mutwo)$ is obtained empirically by
binning the
observational data. Then we construct again the logarithm of likelihood,
considering variations in four parameters: the number of stream stars, the
proper motion of the stream in $\phione$ and $\phitwo$ and the proper motion
spread
$\sigma_\mu$. This likelihood is then maximized  and we determine the
$\muone$, $\mutwo$ and $\sigma_\mu$ for different stream pieces. The right panel
of the
Figure~\ref{fig:obsdata} shows the resulting proper motion estimates as
a function of $\phione$. These proper motions have not been corrected here for
the Sun's reflex motion, which we will model in Section~\ref{sec:orbitfitting}.

Overall, the analysis presented in the previous sections has resulted in the best and
most extensive set of 6-D phase-space coordinate map for a cold stream of stars
in the Milky Way.

\begin{deluxetable}{cccc}
\tablecaption{Radial velocities from the Calar Alto Observations}
\tablehead{ \colhead{Star} &
\colhead{$\phione$} & \colhead{$\phitwo$} & \colhead{$V_{rad}$} \\
 & \colhead{[deg]} & \colhead{[deg]} & \colhead{[km/s]} 
}
\startdata
SDSS J094105.35+315111.6 & $-45.23$ & $-0.04$ & \phs$28.8 \pm 6.9$\\
SDSS J094705.26+332939.8 & $-43.17$ & $-0.09$ & \phs\phn$29.3 \pm 10.2$\\
SDSS J095740.48+362333.0 & $-39.54$ & $-0.07$ & \phs\phn$2.9 \pm 8.7$\\
SDSS J095910.43+363206.6 & $-39.25$ & $-0.22$ & \phn$-5.2 \pm 6.5$\\
SDSS J100222.01+374113.3 & $-37.95$ & \phs$0.00$ & \phs\phn$1.1 \pm 5.6$\\
SDSS J100222.02+374049.2 & $-37.96$ & $-0.00$ & \phn$-11.7 \pm 11.2$\\
SDSS J101033.02+393300.8 & $-35.49$ & $-0.05$ & $-50.4 \pm 5.2$\\
SDSS J101110.08+394453.9 & $-35.27$ & $-0.02$ & \phn$-30.9 \pm 12.8$\\
SDSS J101254.83+395525.6 & $-34.92$ & $-0.15$ & $-35.3 \pm 7.5$\\
SDSS J101312.05+400613.3 & $-34.74$ & $-0.08$ & $-30.9 \pm 9.2$\\
SDSS J101702.15+404747.3 & $-33.74$ & $-0.18$ & $-74.3 \pm 9.8$\\
SDSS J101951.76+412701.5 & $-32.90$ & $-0.15$ & $-71.5 \pm 9.6$\\
SDSS J102216.20+415534.7 & $-32.25$ & $-0.17$ & $-71.5 \pm 9.2$\\
SDSS J103003.87+434351.7 & $-29.95$ & $-0.00$ & $-92.7 \pm 8.7$\\
SDSS J104341.92+460224.7 & $-26.61$ & $-0.11$ & $-114.2 \pm 7.3$\phn\\
SDSS J104840.98+464922.1 & $-25.45$ & $-0.14$ &  $-67.8 \pm 7.1$\\
SDSS J105036.96+472000.1 & $-24.86$ & \phs$0.01$ & $-111.2 \pm 17.8$\\
SDSS J110711.27+494415.9 & $-21.21$ & $-0.02$ & $-144.4 \pm 10.5$\\
SDSS J114242.08+533841.4 & $-14.47$ & $-0.15$ & $-179.0 \pm 10.0$\\
SDSS J114724.59+535546.8 & $-13.73$ & $-0.28$ & $-191.4 \pm 7.5$\phn\\
SDSS J115116.08+542142.7 & $-13.02$ & $-0.21$ & $-162.9 \pm 9.6$\phn\\
SDSS J115326.06+542930.6 & $-12.68$ & $-0.26$ & $-217.2 \pm 10.7$\\
SDSS J115404.06+543511.4 & $-12.55$ & $-0.23$ & $-172.2 \pm 6.6$\phn
\enddata
\label{tab:vels}
\end{deluxetable}

\begin{deluxetable}{cc}
\tablecaption{Stream positions}
\tablehead{\colhead{$\phione$} & \colhead{$\phitwo$} \\
\colhead{[deg]} & \colhead{[deg]} }
\startdata
$-60.00$ & $-0.64 \pm 0.15$\\
$-56.00$ & $-0.89 \pm 0.27$\\
$-54.00$ & $-0.45 \pm 0.15$\\
$-48.00$ & $-0.08 \pm 0.13$\\
$-44.00$ & \phs$0.01 \pm 0.14$\\
$-40.00$ & $-0.00 \pm 0.09$\\
$-36.00$ & \phs$0.04 \pm 0.10$\\
$-34.00$ & \phs$0.06 \pm 0.13$\\
$-32.00$ & \phs$0.04 \pm 0.06$\\
$-30.00$ & \phs$0.08 \pm 0.10$\\
$-28.00$ & \phs$0.03 \pm 0.12$\\
$-24.00$ & \phs$0.06 \pm 0.05$\\
$-22.00$ & \phs$0.06 \pm 0.13$\\
$-18.00$ & $-0.05 \pm 0.11$\\
$-12.00$ & $-0.29 \pm 0.16$\\
\phn$-2.00$ & $-0.87 \pm 0.07$
\enddata
\label{tab:poss}
\end{deluxetable}

\begin{deluxetable}{cc}
\tablecaption{Stream distances}
\tablehead{\colhead{$\phione$} & \colhead{Distance} \\
\colhead{[deg]} & \colhead{[kpc]} }
\startdata
$-55.00$ & $7.20 \pm 0.30$\\
$-45.00$ & $7.59 \pm 0.40$\\
$-35.00$ & $7.83 \pm 0.30$\\
$-25.00$ & $8.69 \pm 0.40$\\
$-15.00$ & $8.91 \pm 0.40$\\
\phs\phn$0.00$ & $9.86 \pm 0.50$
\enddata
\label{tab:distances}
\end{deluxetable}

\begin{deluxetable}{cccc}
\tablecaption{Stream proper motions}
\tablehead{\colhead{$\phione$} & \colhead{$\muone$} & \colhead{$\mutwo$} &
\colhead{$\sigma_{\mu}$} \\
\colhead{[deg]} & \colhead{[mas/yr]} & \colhead{[mas/yr]} &
\colhead{[mas/yr]}}
\startdata
-55.00 & -13.60 & -5.70 & 1.30\\
-45.00 & -13.10 & -3.30 & 0.70\\
-35.00 & -12.20 & -3.10 & 1.00\\
-25.00 & -12.60 & -2.70 & 1.40\\
-15.00 & -10.80 & -2.80 & 1.00
\enddata
\label{tab:pms}
\end{deluxetable}

\section{Orbit fitting}
\label{sec:orbitfitting}

If we can assume that all the stream stars lie close to one single
test-particle orbit,
then our phase space map of the \GS stream 
should not only define this orbit, but at the same time constrain the Milky
Way's potential. 
\change{In general the stars in the tidal stream have slightly
different energies and angular momenta, but the assumption that the stream stars
are moving along the same orbit is plausible , especially if the stream is thin
and is near pericenter \citep[][\& private communication]{dehnen04}.} But it is
not straightforward to quantify the quality of such an approximation. 
For now we simply fit an orbit to our 6-D map of 
available observational data: the
position on the sky, $\phitwo(\phione)$, proper motion $\vec{\mu}(\phione)$,
distance to the stream $D(\phione)$ and radial velocity $V_{rad}(\phione)$. For
each assumed potential, we will determine the best fit orbit, but then
marginalize over the orbits to determine the  range of viable gravitational
potentials.  This analysis extends earlier efforts by 
 \citet{grillmair06_gd} and \citet{willett09} who have presented orbit
solutions for \GS. However, we can now draw on a much more extensive set of
observational constraints. We also explore the
fit degeneracies. \change{Given that our 6-D phase-space map of the \GS stream
spans only
a
limited range in R and z (as seen from the Galactic center
$11$~kpc$\lesssim$R$\lesssim14$~kpc, $5$~kpc$\lesssim$z$\lesssim9$~kpc) it
proved useful
to consider very simple parametrized potentials at first}. Further it proved
necessary to consider what prior information we have on the Sun's
(i.e. the observers) position and motion, as well as on our Milky Way's stellar
disk mass.

\subsection{One component potential}
The stream is located at Galactocentric $(R,z)\approx (12,6)$~kpc,  a regime
where
presumably both the stellar disk and the dark halo contribute to the potential
, and its flattening. Of course, the stream dynamics are solely
determined by the total potential, and therefore we consider first  a simple
single-component potential, the flattened logarithmic potential 
\begin{equation}
\Phi(x,y,z)=\frac{V_c^2}{2}\,
{\rm ln}\left(x^2+y^2+\left(\frac{z}{q_{\Phi}}\right)^2\right)\text{,}
\label{eq:logpot}
\end{equation}
which has only two parameters: the circular velocity $V_c$ and the
flattening $q_\Phi$. Note that $(1-q_{density})\approx 3(1-q_\Phi)$ for
moderate flattening \citep[e.g. p.48 of][]{binney87}. Such a simple potential
seems
justified as the stream stars are only probing a relatively small range in R
and z.

\begin{figure*}
 \includegraphics[width= \textwidth]{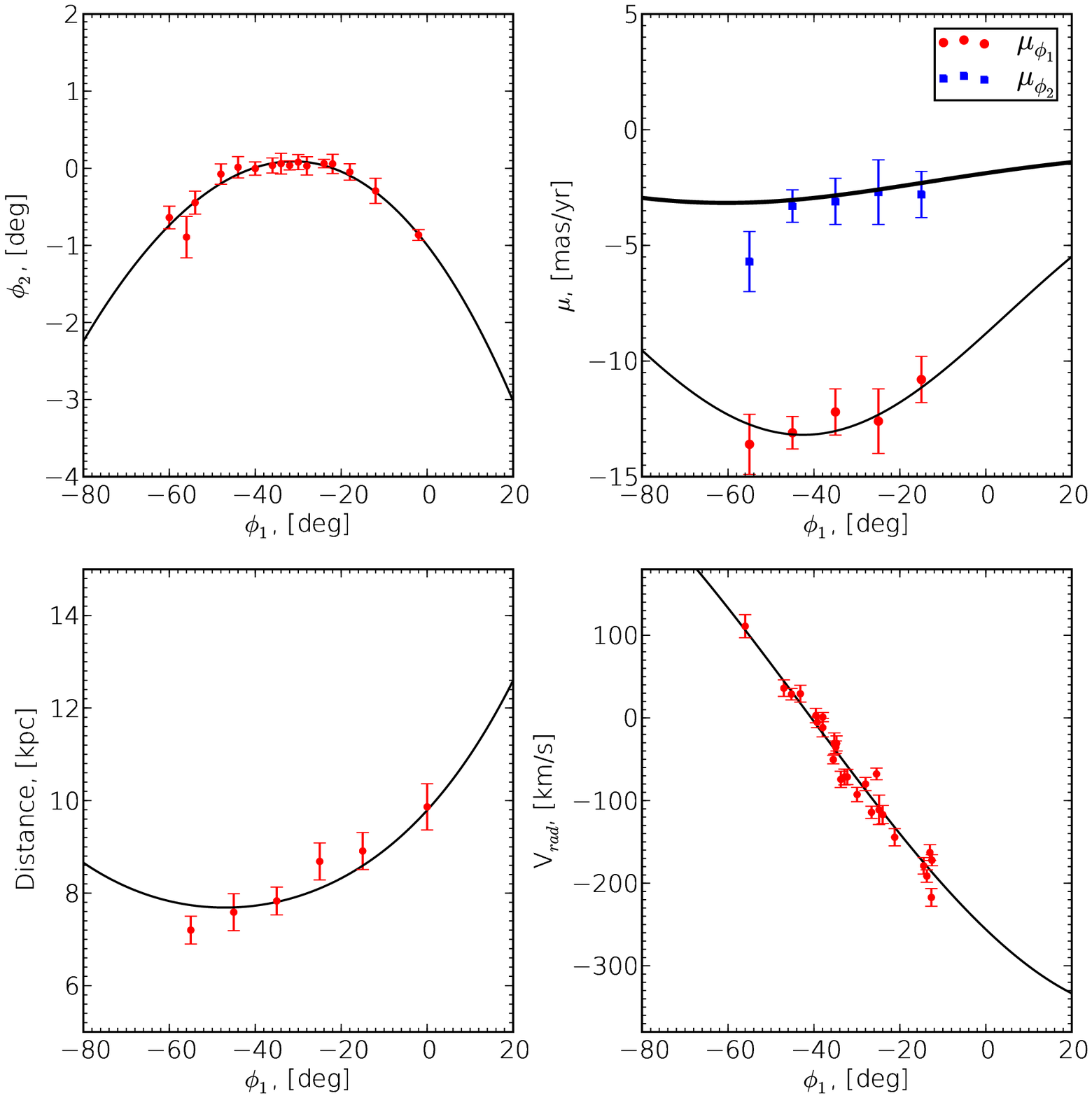}
\caption{The data-model comparison for the best fit orbit in a flattened
logarithmic
potential (Eq.~\ref{eq:logpot} with $V_c=220$~km/s and $q_\Phi$=0.9. The color
data points with error bars shows the observational data, while the black lines
show the model predictions for the orbit with $\vec{X}(0)=(-3.41, 13.00,
9.58)$~kpc, $\dot{\vec{X}}(0)=(-200.4, -162.6, 13.9)$~km/s. The top left
panel shows the positions on the sky, the top right panel shows the
proper motions, the bottom left panel shows the distances, the bottom right
panel shows the radial velocities. On the top right panel, red circles and
thin line show $\muone$, while blue squares and thick line show $\mutwo$.}
\label{fig:orbit_220_0.9}
\end{figure*}

In practice, we fit an orbit to the 6-D stream map by considering a set of
trial starting points in the Galaxy, specified by the initial conditions 
$(\vec{X}(0),\dot{\vec{X}}(0))$ in
standard Cartesian Galactic coordinates. Together with an assumed
gravitational potential this predicts $(\vec{X}(t),\dot{\vec{X}}(t))$, which
can be converted to the observables, 
$\phitwo(\phione)$, $\vec{\mu}(\phione)$, $D(\phione)$, $V_{rad}(\phione)$ and
then compared to the 6-D observations (Fig.~\ref{fig:orbit_220_0.9}).
For each $[(\vec{X}(0),\dot{\vec{X}}(0)|\Phi(\vec{X})]$ we can evaluate the
quality of
the fit by calculating  
$\chi^2$, summing over all data points, shown in
Fig.~\ref{fig:orbit_220_0.9}. For any given
$\Phi(\vec{X})$, $\chi^2$ can be then minimized with respect to the orbit, i.e.
$(\vec{X}(0),\dot{\vec{X}}(0))$, providing the 'best fit' orbit in this
potential and the plausibility of that potential. The minimization is performed
using the MPFIT code \citep{markwardt09} implementing the Levenberg-Marquardt
technique \citep{marquardt63} translated
into Python\footnote{\url{http://code.google.com/astrolibpy/}}.
The data used to constrain the potential are given in the
Tables~\ref{tab:vels}, \ref{tab:poss}, \ref{tab:distances}, \ref{tab:pms}
(except the SDSS measurements of the radial velocities which are given in the
end of Section~\ref{sec:kinematics}).

It is crucial to note that the conversion of $(\vec{X}(t),\dot{\vec{X}}(t))$ to
the space of observables depends on the position and motion of the observer,
i.e. on distance from the Sun to the Galactic center ($R_0$) and on the 3-D
velocity of the Sun in the Galaxy rest-frame ($\vec{V}_0$). At this stage we
adopt $R_0=8.5$~kpc based on recent determinations
\citep[e.g.][]{ghez08,gillessen09}, but
later we will relax this.
The second parameter $\vec{V}_0\equiv \vec{V}_{LSR} + \Delta \vec{V}_{LSR}$
(where $V_{LSR}$ is the velocity of the Local Standard of Rest and $\Delta
\vec{V}_{LSR}$ is the Sun's velocity relative to the LSR) is
linked to
the fitting not only through conversion of the observable relative stream
velocities to the GC rest system, but also conceptually through the plausible
demand that $\Phi(\vec{X})$ and in particular $V_c(R_0,0)$
also reproduces $\vec{V}_{LSR}$. In this way, constraints on the potential
flattening
can be derived by considering $r\frac{d\Phi}{dr}$ in the disk plane
($\vec{V}_{LSR}$)
and the plane of the \GS stream.
The velocity of the Sun relative to the Local Standard of Rest (LSR) $\Delta
\vec{V}_{LSR}$ is quite well known from the HIPPARCOS measurements
\citep{dehnen98}:
$\Delta\vec{V}_{LSR}[km/s]=10\vec{e}_x+5.25\vec{e}_y+7.17\vec{e}_z$.
The velocity of the LSR, i.e. $V_c(R_0,0)$ has a considerable
uncertainty \citep{brand93,ghez08,xue08,reid09}. Initially  we will consider
the velocity of the LSR simply a consequence of the assumed potential, i.e.
${V}_{LSR}\equiv V_c(R_0,0)$.

Figure~\ref{fig:orbit_220_0.9} illustrates the result of such fitting, by
overplotting the best fit
orbit for the plausible potential with $V_c=220$~km/s and $q_\Phi=0.9$ over
observational data. It is clear that even for the simple flattened logarithmic
potential, an orbit can be found that reproduces the observables well. This fit
and Figure serve to illustrate a few generic points that also hold for orbit
fits in differing potentials: the stream moves on a retrograde orbit and
it is near pericenter, where the stream is expected to approximate an orbit
well \citep{dehnen04}. 
After fitting a first orbit, we may also note its global parameters
(see Fig.~\ref{fig:orbit3d} for a 3-D map of the orbit):  pericenter is at
 14~kpc from the GC; apocenter is at 26~kpc; and the inclination
 is $39\degr$.

\begin{figure}
 \includegraphics[width=0.5\textwidth]{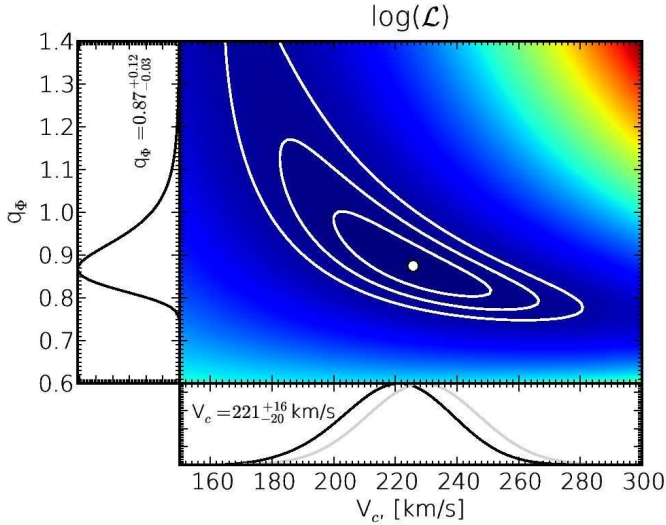}
\caption{The log-likelihood surface of the orbit fit for the family of
flattened logarithmic potentials (Eq.~\ref{eq:logpot}) with different circular
velocities
$V_c$ and flattenings $q_\Phi$ with a flat prior on $V_c$. Note
that $V_c$ enters both into the model velocities of the stream stars and into
the correction of all three velocity components for the Sun's motion.
The contours show the 1$\sigma$, 2$\sigma$ and 3$\sigma$ confidence regions.
The inset panels at the bottom and on the left show the 1D marginalized
posterior probability distributions for $V_c$, $q_\Phi$ respectively. The gray
line in the bottom panel shows the probability distribution for the $V_c$ from
\citet{ghez08}, which we shall use as a prior in
Fig.~\ref{fig:logpot_contours_prior}.}
\label{fig:logpot_contours}
\end{figure}

For any given potential $\Phi(\vec{X}|V_c, q_\Phi)$ we can find best-fit
orbit by marginalizing over $(\vec{X}(0),\dot{\vec{X}}(0))$ to see what our 
6-D map of \GS  implies about the relative plausibility of different  $V_c$ and 
$q_\Phi$: Figure~\ref{fig:logpot_contours} shows the log-likelihood surface
 for the potential parameters $(V_c,q_\Phi)$; note again that this
fit neglects all other prior information on $V_c$ at the Sun's position. The
contours show 1$\sigma$, 2$\sigma$, 3$\sigma$ confidence regions on the
parameters, derived
from the $\delta({\rm ln}(\mathcal{L}))$ values for two degrees of freedom (i.e.
a two parameter fit)\citep[see e.g.][]{lampton76}. The insets at the left and
bottom show the marginalized distributions for single parameters. The best fit
values with the 2-sided 68\% confidence intervals are $V_c=221^{+16}_{-20}$~km/s
and $q_\Phi=0.87^{+0.12}_{-0.03}$. Figure~\ref{fig:logpot_contours} illustrates
that the flattening parameter $q_\Phi$ is quite
covariant with the equatorial circular velocity $V_c$. An extreme example may
serve to explain this covariance qualitatively. If the stream went right over
the pole (z-axis), then the local force gradient would be proportional to
$V_c\times q_\Phi$ (Eq.~\ref{eq:logpot}). Information about the
potential flattening must therefore come from combining kinematics and dynamics
in the disk plane with the information from \GS. 

The fit shown in Fig.~\ref{fig:logpot_contours} asks the data not only
to constrain the potential at the stream location and determine the stream
orbit, but also to infer the Sun's motion (or at least $V_{LSR}$) from its
reflex effect on the data. Clearly providing a prior on $V_c(R_0,0)$ is
sensible, especially if we care about constraints on the potential flattening.
We consider the constraints that arise from the Sun's reflex motion with
respect to the Galactic center the most robust and sensible prior in this
context. \citet{ghez08} recently combined radio data \citep{reid04} with
near-IR data on the Galactic center kinematics to arrive at
$V_c(R_0,0)=229\pm18$~km/s. It is also noticeable that our own constraint on
 $V_c(R_0,0)$ from Fig.~\ref{fig:logpot_contours}, $221^{+16}_{-20}$~km/s, is
close both in value and uncertainty to the estimate of \citet{ghez08}, 
which is based on a completely disjoint dataset and approach.

\begin{figure}
 \includegraphics[width=0.5\textwidth]{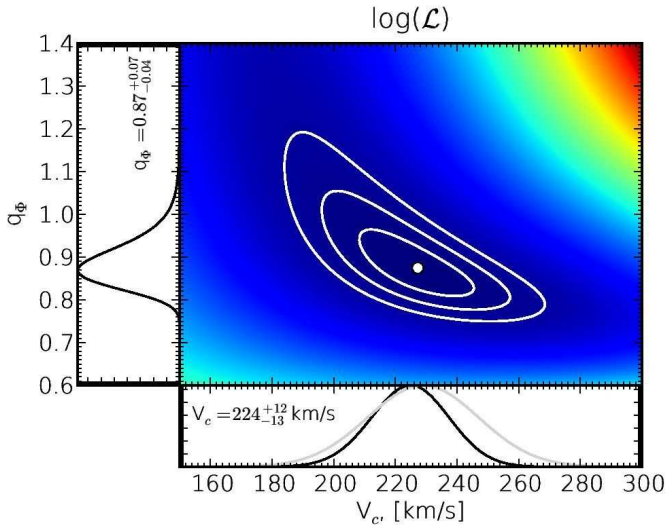}
\caption{The log-likelihood surface of the orbit fit for the family of flattened
logarithmic potentials (Eq.~\ref{eq:logpot}) with different circular velocities
$V_c$ and flattenings $q_\Phi$, but now with a prior on the $V_c$ of
229$\pm$18~km/s from \citet{ghez08}. The likelihood was also marginalized over
the
Gaussian prior on $R_0=8.4\pm0.4$~kpc.
As on Fig.~\ref{fig:logpot_contours} 
the contours show the 1$\sigma$, 2$\sigma$ and 3$\sigma$ confidence regions.
The inset panels at the bottom and on the left show the 1D marginalized
posterior probability distributions for $V_c$, $q_\Phi$ respectively. The gray
line in the bottom panel shows the adopted prior distribution for the $V_c$ from
\citet{ghez08}}
\label{fig:logpot_contours_prior}
\end{figure}

\begin{figure}
 \includegraphics[width=0.5\textwidth]{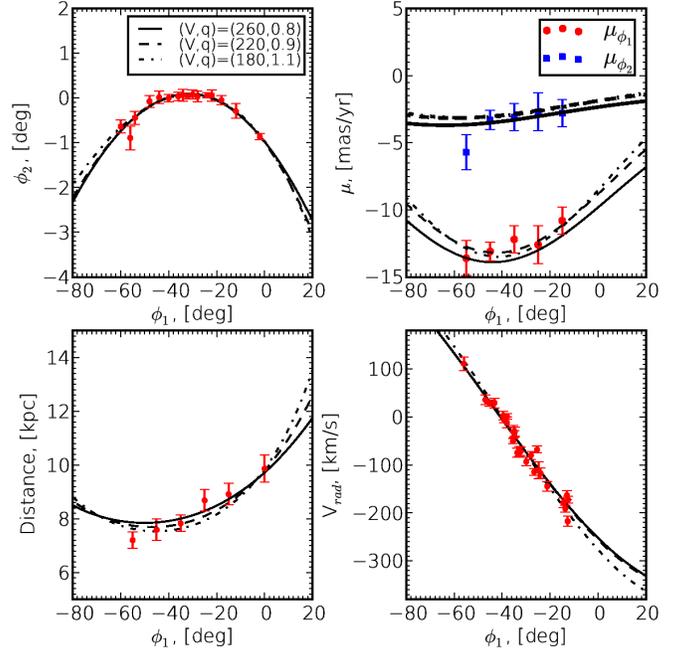}
\caption{The data-model comparison for a set of best-fit orbits in different
logarithmic potentials (Eq.~\ref{eq:logpot}) with three different $(V_c,
q_\Phi)$
 parameters values (180~km/s,1.1), (220~km/s,0.9), (260~km/s,0.8.) The colored
data points with error bars show the observational data, while the black lines
show the model predictions (different line styles show the orbit models in
different potentials). The top left
panel shows the positions on the sky, the top right panel shows the
proper motions, the bottom left panel shows the distances, the bottom right
panel shows the radial velocities. On the top right panel, red circles and
thin lines show $\muone$, while blue squares and thick lines show $\mutwo$.}
\label{fig:orbit_3models}
\end{figure}

Fig.~\ref{fig:logpot_contours_prior} shows the resulting log-likelihood
contours and 1$\sigma$, 2$\sigma$, 3$\sigma$ confidence regions after applying
this prior on
the $V_c$ (229$\pm$18~km/s). Note that likelihood
on Fig.~\ref{fig:logpot_contours_prior} is also marginalized over $R_0$ with
a Gaussian prior \citep[$R_0=8.4\pm0.4$~kpc][]{ghez08}.

Fig.~\ref{fig:logpot_contours_prior} shows that the posterior probability
distribution on $V_c$ has slightly changed to $V_c(R_0)=224^{+12}_{-14}$~km/s
with noticeably smaller error bar compared to the value from \citet{ghez08}.
Fig~\ref{fig:logpot_contours_prior} \change{also shows us the slight improvement
comparing to Fig.~\ref{fig:logpot_contours} of flattening constraints:}
$q_\Phi=0.87^{+0.07}_{-0.04}$.
This means that the {\it total} potential appears to be oblate (in the radial
range probed); this may not be surprising, as the stellar
disk -- which is manifestly very flattened -- contributes to the total
potential.

Fig.~\ref{fig:orbit_3models} illustrates how well the best-fit
orbits for different potentials $\Phi(\vec{X}|V_c,q_\Phi)$ can mimic one another
other in the space of observables. This is the source of the parameter
covariances shown in Fig.~\ref{fig:logpot_contours}
and \ref{fig:logpot_contours_prior}.

The fitting of the orbit shown in Fig.~\ref{fig:orbit_220_0.9} allows us to make
an estimate of the line-of-sight velocity dispersion in the stream, by comparing
the dispersion of the radial velocity residuals with the accuracy
of individual radial velocities. \change{Assuming that without observational
uncertainties the residuals from the orbit fit should be Gaussian distributed
with zero mean we performed the maximum likelihood fitting of the residuals, 
which gives a 90\% confidence upper limit on the velocity dispersion of stars
in the stream of $\sim 3$~km/s.}

Before we aim at separating possible flattening contributions from
the halo and disk, it is worth commenting on the accuracy and limitations of our
estimate of $q_\Phi$. In the range $(R,z) \approx (12,6)$~kpc no
other direct observational constraints on the potential shape exist in the
literature, and hence our estimate of $q_\Phi=0.87^{+0.07}_{-0.04}$ is a new and
important contribution. On the other hand, an error of  $\delta q_\Phi \sim
0.05$, especially when translated into the flattening error of
the equivalent scale-free mass distribution, may not appear as particularly
helpful in model discrimination, or as impressively accurate. Especially as a
manifestly cold stellar stream spanning over $60\degr$ on the sky may seem ideal
for mapping the potential at first glance.

\begin{figure}
 \includegraphics[width=0.5\textwidth]{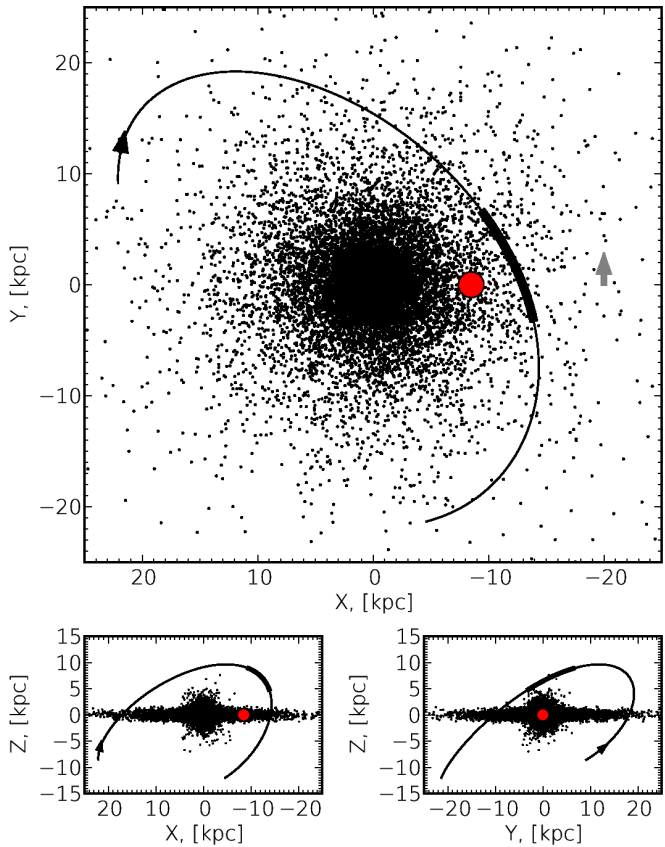}
\caption{2-D projections of the orbit in the Galactic rectangular coordinates.
The position of the Sun is shown by a red circle. The Galaxy is shown by a
cloud of points and the gray arrow shows the direction of the galactic rotation.
The black arrow shows the direction of the orbital movement of the stream stars.
The orbit is the best fit orbit for the $V_c=220$~km/s, $q_\Phi=0.9$ logarithmic
potential(Eq.~\ref{eq:logpot}). The orbit for the 3-component
potential(Eqns.~\ref{eq:halopot}, \ref{eq:diskpot}, \ref{eq:bulgepot}) is almost
undistinguishable from the orbit in logarithmic potential.}
\label{fig:orbit3d}
\end{figure}

\subsection{Constraints on the shape of the dark matter halo from a bulge,
disk, halo 3-component potential}

In the previous section we constrained the parameters of a simplified MW
potential, the spheroidal logarithmic potential.
It is clear that the MW potential at the position of the
stream must depend explicitly on the sum of baryonic Galaxy components
(bulge and disk) and on the dark matter halo. We now explore whether our
constraint on the shape of the overall potential, $q_\Phi \sim 0.9$, permits
interesting statements about the shape of the DM potential itself. At the
distance of $(R,z)\approx (12,6)$~kpc the
contribution of the disk to the potential should still be relatively
large, weakening or at least complicating inferences on the shape of
the DM distribution.

We adopt a three-component model of the Galaxy potential, choosing one that is 
widely used  in the modeling of the Sgr stream
\citep{helmi04,law05,fellhauer06} and reproduces the galactic
rotation curve reasonably well.

The model consists of a halo, represented by the logarithmic potential
\change{The numerical values of parameters of the potential were taken from 
\citet{fellhauer06}}
\begin{equation}
\Phi_{halo}(x,y,z)=\frac{v_{halo}^2}{2}\,
{\rm ln}\left(x^2+y^2+\left(\frac{z}{q_{\Phi,halo}}\right)^2+d^2\right)\text{,}
\label{eq:halopot}
\end{equation}
where we have adopted  $d=12$~kpc from the previous authors. The disk is
represented by a Miyamoto-Nagai potential \citep{miyamoto75},
\begin{equation}
\Phi_{disk}(x,y,z)=\frac{G M_{disk}}{\sqrt{x^2+y^2+(b+\sqrt{z^2+c^2})^2}}
\label{eq:diskpot}
\end{equation}
with $b=6.5$~kpc, $c=0.26$~kpc. The bulge is modeled as a Hernquist potential:
\begin{equation}
\Phi_{bulge}(x,y,z)=\frac{G M_b}{r+a}
\label{eq:bulgepot}
\end{equation}
with $M_b=3.4\times10^{10} M_\sun$, $a=0.7$~kpc

\begin{figure}
 \includegraphics[width=0.5\textwidth]{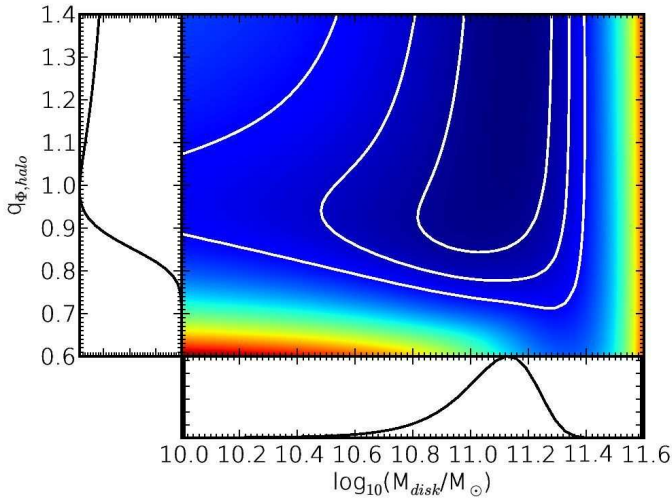}
\caption{The log-likelihood surface of the orbit fit for a 3-component
potential (Eq.~\ref{eq:halopot}, \ref{eq:diskpot} and \ref{eq:bulgepot})  with
different disk masses $M_{disk}$, halo circular velocities $v_{halo}$
and halo flattenings $q_{\Phi,halo}$. The likelihood was marginalized over the
circular velocity of the halo $v_{halo}$ with the Gaussian prior
on $v_{halo}=170\pm15$~km/s from \citet{xue08} and the Gaussian prior on
$V_{LSR}=229\pm 18$~km/s) from \citet{ghez08} . The contours show the
1$\sigma$, 2$\sigma$ and 3$\sigma$ confidence regions.
The inset panels at the bottom and on the left show the 1D marginalized
posterior probability distributions for $M_{disk}$, $q_{\Phi,halo}$
respectively.}
\label{fig:3comppot}
\end{figure}

As in the previous section, for any given set of potential parameters,
we can find the best-fitting stream orbit and compute $\chi^2$
of the fit. In the current paper we do not make any attempts to fully fit all
the parameters of the MW  potential, but we try to make an estimate of the MW DM
halo flattening.
We take the 3-component potential and fix all but 3
parameters --- disk mass $M_{disk}$, circular velocity of the
halo $v_{halo}$ and the flattening of the halo $q_{\Phi,halo}$. On a 
3-D grid of
these parameters we perform a $\chi^2$ fit. Figure~\ref{fig:3comppot}
shows the results of such fit after marginalization over
the orbital parameters $(\vec{X}(0),\dot{\vec{X}}(0))$, circular
velocity of the halo $v_{halo}$ with a Gaussian prior from
\citet{xue08} and a Gaussian prior on the circular velocity at the Sun's
radius from \citet{ghez08}. The Figure clearly illustrates that in the case of
the 3-component potential, the current data is unable to give a significant new
insights on the flattening of the DM halo. We can only say that at 90\%
confidence $q_{\Phi,halo}>0.89$. 
We note however that for the future analysis, if a multi-component model
for the potential is used then more prior information is required, i.e. not
only on $V_{LSR}$ and $v_{halo}$ but on $M_{disk}$ and other parameters of the
potential.

\section{Discussion and Conclusions}

In this paper we have presented a thorough analysis the \GS stream combining the
publicly available SDSS and SEGUE data with follow-up spectroscopic
observations from Calar Alto. The combination of the photometric SDSS
observations, USNO-B/SDSS proper motions, SDSS, SEGUE and Calar Alto radial
velocities allowed us to construct a unprecedented 6-D phase-space map of this
very faint (29 mag/sq. arcsec) tidal stream. The 6-D phase-space map of the
stream, spanning more than $60\degr$ on the
sky, provided the opportunity not only to fit the orbit as
\citet{grillmair06_gd}
and \citet{willett09} have done previously but to explore what 
constraints can be placed on the MW potential. 

The analysis is based on the assumption that the stream stars occupy one orbit.
In detail, of course, different stars on the stream have slightly different
values of conserved quantities and therefore lie on slightly different orbits.
Effectively, our analysis depends on these being small when compared with an
orbital uncertainties. The magnitude of the departure of the stream from a
single orbit will, in detail, be a function of the projenitor and the
distruption process; as these details became understood the model can be
refined.

We found that the distribution of stream stars in phase space can
be well fit by an inclined eccentric orbit in the spheroidal logarithmic
potential. After marginalization over the stream orbital parameters we
derive a circular velocity $V_c=221^{+16}_{-20}$~km/s and flattening
$q_\Phi=0.87^{+0.12}_{-0.03}$. This measurement has been made without the use
of any information other than that in the \GS stream itself. It is important
that the
information available in the observations of the stream is very sensitive to the
$V_{c}$, the circular velocity at the Sun's position. The reason for that is
that the stream extends more then 60$\degr$ on the sky and therefore both the
radial velocities and the proper motions have components coming from the
projection of the Sun's motion.

If we combine our circular velocity measurement with existing prior on the
$V_c$  from \citet{ghez08} and also marginalize over the distance from the Sun
to the Galactic center using the \citet{ghez08} prior ($R_0=8.4\pm0.4$~kpc), we
further tighten the error bar on $V_{c}=224^{+12}_{-13}$~km/s and on the
flattening of the potential $q_\Phi=0.87^{+0.07}_{-0.04}$. Our measurement of
the $V_c$ is the best constraint to date on the circular velocity at the Sun's
position, and the measurement of $q_\Phi$ is the only strong constraint on
$q_\Phi$ at galactocentric radii near $R \sim 15$~kpc.

The measurement of the flattening of the potential
$q_\Phi=0.87^{+0.07}_{-0.04}$ describes only the flattening of the overall
Galaxy potential at the stream's position $(R,z) \approx (12,6)$~kpc where the
disk contribution to the potential is presumably large. 
Unfortunately the data on the \GS stream combined with the \citet{ghez08},
\citet{xue08} priors on $V_c$ and $v_{halo}$ are not enough for separating
the flattening
of the halo from the flattening of the total Galaxy potential. So we are unable
to place strong constraints on flattening of the MW DM halo; we put a
90\% confidence lower limit at $q_{\Phi,halo}>0.89$.

Despite the negative result on the measurement of the MW DM halo flattening, we
note that the data from the \GS stream is able
to give strong constraints on two important Galaxy parameters. We claim that
that our dataset on the \GS stream  should now be combined \change{with other
available MW kinematical data (from other stellar streams,
BHBs, MW rotation curve)} in
order to tighten the existing constraints on the Galaxy
parameters. It is important that the constraints on Galaxy potential based
on the \GS stream dataset are, to large extent, model-independent and purely
kinematic i.e. the constraints on the $V_{c}$ come to large extent from the
projection effects and manifest themselves in proper motions and radial
velocities.

\change{We think that it is quite surprising that such a long (60\degr) stream
with full 6-D map didn't allow us to contrain large number of parameters of the
MW DM halo and other Galactic components. We think that there are several
reasons for that, but the most important is that while the observed part of the
orbit spans $\sim$ 70$\degr$ on the sky from the Sun's point of view, the
orbital phase spanned by the stars as seen from the Galactic center, is only
$\sim 40\degr$. Furthermore, the observable part of the stream occupies the
perigalacticon part of the orbit, so the range of Galactocentric distances
probed by the stream is small.}
That makes it plausible why orbits of different eccentricities,
and hence of different azimuthal velocities at their pericenter, can match so
closely the same set of 6-D coordinates. Since very cold streams take many orbits
to spread a substantial fraction of 2$\pi$ in
orbital phase \citep[see e.g. also Pal 5;][]{odenkirchen01,grillmair06_pal5},
all future analyses of yet-to-be discovered streams will particularly 
need to consider the trade-off between the conceptual and practical
attractiveness of 'cold' streams and the near-inevitable limitations of 
their phase coverage.

In addition to the weakness coming from phase coverage, our analysis at
this point must rely on photometric distance estimates; these have random errors
of $\sim$10\%, after an empirical distance correction to the best fit isochrones
that is of the same magnitude (see Section~\ref{sec:stellarpop}). While the
proper motions that we derived for ensembles of stream stars are unprecedented
for a stellar stream in the Milky Way's outer halo; yet, the corresponding
velocity precision, especially when compounded by distance errors, is still the
largest single source of uncertainty in the fitting (e.g. our tests have
shown that overestimating heliocentric distances to the stream leads to the
overestimated measurement of $V_c$). \change{Both the deficiencies of the
distances and proper motions on \GS will be largely alleviated after the launch
of the Gaia satellite, which will allow much tighter constraints on the
Galactic potential.}

\change{But before the launch of Gaia, we suggest that further observations of
radial velocities of stars in the stream (with deeper spectroscopy) and
improvements in proper motion precision \citep[with e.g. Pan-STARRS][]{kaiser02}
should be able to reduce the error bars on Galaxy parameters significantly. It
is also important to calibrate properly the distance to the stream, which may be
done by confirming several probable BHB candidates in the stream. }

We suggest that any attempt to fit the Galactic potential\citep[such
as][]{widrow08} now should not ignore the dataset on \GS and should incorporate
it into their fits.

The orbital parameters, which  we have measured for the \GS stream, are more or
less consistent with those from \citet{willett09}: pericenter is at 14~kpc from
the GC, the apocenter is at 26~kpc, the orbit inclination with respect to the
Galactic plane is $39\degr$. \change{ We have also estimated the total stellar
mass associated with the stream to be $\sim 2\times  10^4M_\odot$, which
together with the relatively small stream width of $\sim 20$~pc suggests that
the progenitor of the stream was a globular cluster, although we cannot
completely rule out the dwarf galaxy progenitor. We have also determined the
90\% confidence upper limit for the velocity dispersion of the stars in the
stream $\sim 3$~km/s which do not contradict significantly neither globular
cluster nor dwarf galaxy progenitor hypothesis. Given the length of the
observable part of the stream of $\sim 10$~kpc, and the velocity dispersion
$\lesssim3$~km/s the age of the stream can be estimated to be greater than
$1.5$~Gyrs (assuming that the progenitor is in the middle of the observed part
of the stream). It is interesting that the stream managed to evade possible
destruction by interaction with DM subhalos orbiting around
MW \citep{carlberg09}. Although the clumpiness observed in the stream may be
attributed to these past interactions (Koposov et al 2010 in prep.).}

Overall in this paper we have illustrated a method of analyzing the thin
stellar stream using all the available information on it and further utilizing
that to constrain the Galaxy potential. We believe that in the epoch of
Pan-STARRS, LSST and especially Gaia, which will give us a wealth of new
information on the MW halo, stream-fitting like that presented in
this paper will be extremely useful and productive.

\acknowledgments{
SK was supported by the DFG through SFB 439, by a EARA-EST Marie
Curie Visiting fellowship and partially by RFBR 08-02-00381-a grant. SK
acknowledges hospitality from the Kavli Institute for Theoretical Physics (KITP)
Santa Barbara during the workshop ``Building the
Milky Way''. SK thanks Jelte de Jong for running MATCH code on the
\GS data and the Calar Alto observing staff for excellent support. DWH
acknowledges support from NASA (grant NNX08AJ48G) and a Research Fellowship from
the Alexander von Humboldt Foundation. The authors thank Kathryn Johnston,
Jorge Pe\~{n}arubia, James Binney for
useful discussions and the anonymous referee for the elaborated referee report
which helped us improve the paper.

The project made use the open-source Python modules matplotlib, numpy, scipy;
SAI Catalogue Access Services (Sternberg Astronomical Institute, Moscow,
Russia)\citep{koposov07_saicas}; NASA's Astrophysics Data System Bibliographic
Services.

The SDSS is managed by the Astrophysical Research Consortium for the
Participating Institutions. The Participating Institutions are the American
Museum of Natural History, Astrophysical Institute Potsdam, University of Basel,
University of Cambridge, Case Western Reserve University, University of Chicago,
Drexel University, Fermilab, the Institute for Advanced Study, the Japan
Participation Group, Johns Hopkins University, the Joint Institute for Nuclear
Astrophysics, the Kavli Institute for Particle Astrophysics and Cosmology, the
Korean Scientist Group, the Chinese Academy of Sciences (LAMOST), Los Alamos
National Laboratory, the Max-Planck-Institute for Astronomy (MPIA), the
Max-Planck-Institute for Astrophysics (MPA), New Mexico State University, Ohio
State University, University of Pittsburgh, University of Portsmouth, Princeton
University, the United States Naval Observatory, and the University of
Washington.
}
\appendix
\section{Transformation of equatorial coordinates to stream coordinates
$\phione,\phitwo$}
$$\begin{pmatrix}
{\rm cos}(\phione)\,{\rm cos}(\phitwo)\\
{\rm sin}(\phione)\,{\rm cos}(\phitwo)\\
{\rm sin}(\phitwo)\\
\end{pmatrix}=
\begin{pmatrix}
-0.4776303088 & -0.1738432154 & 0.8611897727 \\
 0.510844589  & -0.8524449229 & 0.111245042 \\
 0.7147776536 &  0.4930681392 & 0.4959603976
\end{pmatrix}\times
\begin{pmatrix}{\rm cos}(\alpha)\,{\rm cos}(\delta)\\
{\rm sin}(\alpha)\,{\rm cos}(\delta)\\
{\rm sin}(\delta)\\
\end{pmatrix}
$$

\end{document}